\documentclass{aa}  
\usepackage{graphicx}
\usepackage{txfonts}
\usepackage{gensymb}
\usepackage{multirow}
\usepackage{amsmath}
\usepackage{comment}
\usepackage{tabularx}

\usepackage{natbib}
\usepackage[]{hyperref} 
\usepackage[rightcaption]{sidecap}

\begin{document}

\title{Constraints on the magnetic field in the inter-cluster bridge A399--A401}
\titlerunning{Magnetic field in the A399--A401 bridge}
\authorrunning{Balboni et al.}
\author{
$\rm{M.~Balboni^{1,2},~ A.~Bonafede^{3,4},~ G.~Bernardi^{4,5,6}, ~D. ~Wittor^{7,3}, ~F. ~Vazza^{3,4}, A.~ Botteon^{4}, ~E. ~Carretti^{4},}$ $\rm{ ~T.~ Shimwell^{9}, ~V.~ Vacca^{10}~and~ R.~J.~van~Weeren^{8}}$
      }
\institute{
$^{1}$ INAF - IASF Milano, via A. Corti 12, 20133 Milano, Italy\\
$^{2}$ DiSAT, Universit\`a degli Studi dell’Insubria, via Valleggio 11, I-22100 Como, Italy\\
$^{3}$ DIFA - Universit\`a di Bologna, via Gobetti 93/2, I-40129 Bologna, Italy\\
$^{4}$ INAF - IRA, Via Gobetti 101, I-40129 Bologna, Italy\\
$^{5}$ Department of Physics and Electronics, Rhodes University, PO Box 94, Makhanda, 6140, South Africa\\
$^{6}$ South African Radio Astronomy Observatory (SARAO), Black River Park, 2 Fir Street, Observatory, Cape Town, 7925, South Africa\\
$^{7}$ Hamburger Sternwarte, Gojenbergsweg 112, 21029 Hamburg, Germany\\
$^{8}$ Leiden Observatory, Leiden University, PO Box 9513, 2300 RA Leiden, The Netherlands\\
$^{9}$ ASTRON, the Netherlands Institute for Radio Astronomy, Postbus 2, 7990 AA, Dwingeloo, The Netherlands\\
$^{10}$ INAF-Osservatorio Astronomico di Cagliari, Via della Scienza 5, I-09047 Selargius (CA), Italy}
%
%
\abstract{
Galaxy cluster mergers are natural consequences of the structure formation in the Universe. Such events involve a large amount of energy ($\sim 10^{63}$~erg) dissipated during the process. Part of this energy can be channelled in particle acceleration and magnetic field amplification, enhancing non-thermal emission of the intra- and inter-cluster environment. Recently, low-frequency observations have detected a bridge of diffuse synchrotron emission connecting two merging galaxy clusters, Abell 399 and Abell 401. Such a result provides clear observational evidence of relativistic particles and magnetic fields in-between clusters.
In this work, we have used LOw Frequency ARray (LOFAR) observations at 144 MHz to study for the first time the polarised emission in the A399-A401 bridge region. No polarised emission was detected from the bridge region. 
Assuming a model where polarisation is generated by multiple shocks, depolarisation can be due to Faraday dispersion in the foreground medium with respect to the shocks.
We constrained its Faraday dispersion to be greater than 0.10~rad~m$^{-2}$ at 95\% confidence level, which corresponds to an average magnetic field of the bridge region greater than 0.46~nG (or 0.41~nG if we include regions of the Faraday spectrum that are contaminated by Galactic emission). This result is largely consistent with the predictions from numerical simulations for Mpc regions where the gas density is $\sim 300$ times larger than the mean gas density.}
\keywords{
magnetic fields -- galaxy clusters -- A399-A401 -- polarisation
}
\maketitle
%
%
\section{Introduction}\label{Sec:Intro}

Galaxy clusters are the most massive, gravitationally-bound structures
in the Universe. They are the natural outcome of the hierarchical process of structure formation, where clusters grow through events of merging and accretion of small substructures. 
This accretion process occurs inside the so-called Cosmic Web, made of elongated filaments of matter (galaxies, dark matter and magnetised gas called Warm Hot Intergalactic Medium, WHIM) located between clusters and through which matter flows and collapses onto such objects.

In the past decades, non-thermal, diffuse radio emission has been widely observed in galaxy clusters,
implying particle re-acceleration on clusters scales
\citep[e.g.][for a recent observational review]{vanWeeren19}. Merger events could provide the energy necessary for particle re-acceleration \citep[e.g.][for a theoretical review]{Brunetti2014}.
The most striking examples of these processes in clusters are giant radio halos and relics, which have been widely used to study ICM inner regions. The new generation of radio-telescopes is unveiling that cluster diffuse radio emission is more extended than previously reported \citep[e.g.][]{Kamlesh2021-j0717,Kamlesh2021-A2744,Botteon2022,Cuciti2022}, showing its presence even outside the cluster regions in form of bridges \citep[e.g.][]{Govoni19, Botteon2020, Venturi2022}.
The origin of the observed ($\sim \rm{\mu G}$) magnetic fields that accelerate particles within clusters and beyond remains largely uncertain. A commonly accepted hypothesis is that they result from the amplification of much weaker, pre-existing seeds via shock/compression and/or turbulence/dynamo mechanisms during merger events and structure formation. Therefore, magnetic fields will emerge with different intensities at different physical scales as the result of turbulent motions \citep[see][for reviews on magnetic field amplification at cluster scales]{Donnert2018, Vazza2021}.
The origin of seed fields can either be primordial, i.e. generated in the early Universe prior to recombination, or produced locally at later epochs of the  Universe, in early stars and/or (proto)galaxies, and then injected in the interstellar and intergalactic medium \citep[e.g.,][for reviews]{Widrow2012, Subramanian2016}. Another source of magnetic field seeds can be the feedback events following the gas cooling and the formation of first structures, such as stellar populations or black holes. These astrophysical sources can inject magnetic fields inside circumgalactic medium and cosmic voids at $z \leq 10$ in an inside-out scenario from galaxies to larger scales \citep[e.g.][]{Vazza2021}.
A possible discriminant among models would be the estimate of the magnetic field strength in rarefied environments, like filaments, sheets and voids. Simulations show, indeed, that magnetic field profiles resulting from different models tend to diverge beyond the periphery of galaxy clusters, due to the model-dependent efficiency in producing large-scale magnetic field \citep{Vazza2017}.
Despite the difficulties to observe such faint filamentary emission outside clusters at radio wavelengths, previous works have already constrained the magnetic field strength (up to a few tens of nG) on scales larger than typical cluster sizes ($\gtrsim$ Mpc)
\citep{Neronov2010, Pshirkov16, Brown17, Vernstrom17, Paoletti19, Natwariya21}.     
Different observational approaches have been identified in order to study large-scale magnetic fields. A promising one is the Faraday rotation technique \citep[e.g.,][]{Govoni04}, which estimates the line of sight magnetic field.
Exploiting this effect, \cite{Vernstrom19} studied the emission from extragalactic sources and placed an upper limit on the cosmic magnetic field strength at 40~nG. Similarly, \cite{OSullivan2020} used an analogous technique to study prospectic or real pairs of sources, obtaining upper limits on the co-moving cosmological magnetic field ($B_0$) of $B_0 \leq 4$~nG on Mpc scales. 
\cite{Carretti2022a} recently used $\sim 144$ MHz observations of several filaments to study Faraday rotation properties of such low-density regions and how they evolve with redshift, deriving an average magnetic field $B_{f} = 32 \pm 3$~nG in filaments. In a follow-up work, \cite{Carretti2022b} compared LOFAR observations with MHD cosmological simulations and found that the magnetic field in cosmic web filaments at $z = 0$ is in the $8 - 26 ~ \rm{nG}$ range for a typical filament gas overdensity\footnote{We define the overdensity as the ration between the density and its mean at a given redshift: $\delta (z) = \frac{\rho (z)}{\langle \rho (z) \rangle}$} of $\delta_g=10$.
\cite{Vernstrom2021} used pairs of luminous red galaxies as a tracer of cluster pairs. Using a stacking technique, they estimated the intensity of the magnetic field on $1 - 15$~Mpc scales to be in the $30 - 60$~nG range.
This result implied that primordial magnetic field seeds should be more than a factor of $\sim 6$ higher than the simulated ones using only shock acceleration \citep[see also][]{Hodgson2021}. With a different approach (injection), \cite{Locatelli2021} combined upper limits on the radio emission from two inter-cluster filaments and numerical simulations of the magnetic cosmic web, to constrain the intergalactic magnetic field in the $0.2 - 0.6 \mu $G on 10~Mpc scales. Even more recently, \citep{Vernstrom2023} used stacking to derive information on polarised emission, and magnetic field, in the clusters' peripheries and inter-cluster filaments. They found a diffuse polarised emission with a $\sim 20\%$ polarised fraction, which can be attributed to a Fermi-type re-acceleration process as a consequence of large scale accretion, implying, also, an ordered magnetic field.\\
Recent low frequency ($\sim 140$~MHz) observations with LOFAR have detected diffuse synchrotron emission (a radio bridge) connecting the two merging galaxy clusters, Abell~399 and Abell~401 \citep{Govoni19}, providing observational evidence of relativistic particles and magnetic fields in the inter-cluster region. The origin of the non-thermal emission is still uncertain, although it could be caused by multiple weak shocks present in the bridge region that re-accelerate a pre-existing population of mildly relativistic particles \citep{Govoni19}. Alternatively, turbulent reacceleration in the early phases of the merging event could amplify magnetic fields and reaccelerate particles \cite{BrunettiVazza20}. 
In this work, we have used new LOFAR observations to study the polarised emission of the A399--A401 radio bridge, and provide constraints on the magnetic field in the bridge.
The paper is organized as follows: in Section~\ref{sec:a399_pair}, we describe the target, its properties and past studies; in Section~\ref{sec:data_analysis}, we present the observations, data reduction and imaging process used in this work; in Section~\ref{sec:project} we constrain the Faraday dispersion of the depolarisation mechanism that we use to constrain the average intensity of the magnetic field in Section~\ref{sec:constraints_mag_field}. We conclude in Section~\ref{sec:conclusions}.
Following \cite{Nunhokee2023}, we used the Planck cosmology throughout our work \citep{Planck20}, where $1" = 1.345$~kpc at the cluster pair distance.


\section{The A399--A401 pair}
\label{sec:a399_pair}
A399 and A401 are two massive galaxy clusters separated by a projected distance of $\sim 3$~Mpc, 
both hosting a radio halo \citep{Murgia2010}, likely powered by single objects matter accretion.
Radio, optical and X-ray observations support the scenario where the system is in the initial phase of merger and the two clusters have not yet started to interact \citep[e.g.,][]{Fujita96, Sakelliou04, Murgia2010, Govoni19}.
X-ray observations show the presence of a hot gas ($\sim 7-8$~keV) not only inside the central parts of the two objects but also in the connecting inter-cluster region ($6.5$~keV), which shows enhanced X-ray emission \citep{Akamasu2017}. Moreover, observations of the Sunyaev-Zeldovich (SZ, \citealt{SZ1969}) effect confirmed the presence of a connecting bridge between the two clusters \citep{Planck2013, Bonjean2018, Hincks2021-bis}.
\begin{table}
\small
\rule[0.1cm]{\linewidth}{0.5mm}\\
\begin{tabular}{c|ccccc}
     Object & $z$ & RA & DEC & Mass  & \\ 
     & & & & ($10^{14} M_{\odot}$) & \\ 
     \hline
     Abell 0399 & 0.0718 & $02^{\rm {h}} \ 57^{\rm{m}} \ 56^{\rm s}$  & $+13 
     \degree \ 00' \ 59"$ & $5.7$ & \\
     Abell 0401 & 0.0737 & $02^{\rm h} \ 58^{\rm m} \ 57^{\rm s}$  & $+13\degree \ 34' \ 46"$ & $9.3$ \\
\end{tabular}
\rule[0.1cm]{\linewidth}{0.5mm}
\caption{\normalsize \label{tab:Tabella-1} Summary of the main characteristics of the A399--A401 pair.}
\end{table}
Information about the system are summarised in Tab.~\ref{tab:Tabella-1}.\\
\cite{Govoni19} studied the A399--A401 cluster pair, detecting both diffuse and compact radio emission. 
They also discovered radio emission connecting the two radio halos, providing evidence of relativistic electrons and magnetic fields on Mpc scales in the inter-cluster environment.
\cite{Nunhokee2023} used observations at 346~MHz, where they did not detect the bridge, in order to constrain the average bridge spectral index to be $\alpha > 1.5$ at 95\% confidence level.
Even more recently \cite{deJong2022} and \cite{Radiconi2022} investigated the thermal and non-thermal inter-cluster emission of the pair. \cite{Radiconi2022} did not find any correlation between radio and X-ray emission. Instead, with deeper observations and an improved calibration method, \cite{deJong2022} found a positive trend between the thermal and non-thermal emission in the bridge.
Although the spectral index constraints disfavour the scenario where particles are accelerated by weak shocks, no stringent conclusions about the emission model and no constraints on the magnetic field in the bridge region have been obtained so far.

\section{Observations and data analysis}
\label{sec:data_analysis}

Observations used in this work are part of the LOFAR Two-meter Sky Survey \citep[LoTSS,][]{LoTSS2017, LoTSS-DR1, Shimwell2022}, a deep, $120 - 168$~MHz survey of the Northern sky with a $\sim 100$~$\mu$Jy~beam$^{-1}$ sensitivity at $\sim 6''$ resolution.
The A399--A401 pair was observed within a single LoTSS pointing for 8~h, divided into two sets of 4~h each, due to its low declination ($\delta \sim 13 \degree$, see Tab.~\ref{tab:observation}). Each 4~h set was followed by 10 minutes of calibrator observations, 3C196. As we are performing polarisation studies, the direct combination of both observations would likely introduce depolarisation if not properly corrected for the likely different polarisation angle. For this reason, we only used one of the two tracks, n. L576817.
\begin{table*}
    \begin{tabular}{ ccc|cccc } 
    Pointing & RA & DEC & Obs ID & Frequency & On source time
    & RMS at $20''$ ($\rm{\mu Jy~beam^{-1}}$) \\
    \hline
    \multirow{2}{3.5em}{P043+14} & \multirow{3}{5em}{$02^{\rm h} \ 58^{\rm m} \ 22^{\rm s}$ }& \multirow{2}{5em}{$+13 \degree \ 20' \ 22''$} &  L576817 & 120-168 MHz & 4 hrs  & 445\\
    & & & L573953 & 120-168 MHz & 4 hrs  & 443
    \end{tabular}
    \caption{Description of the available observations.}
\label{tab:observation}
\end{table*}
We used the visibilities calibrated by the \texttt{PREFACTOR3} pipeline \citep{PREFACTOR2019} and by the direction-independent calibration in the LoTSS-DR2 pipeline \citep{Shimwell2022}. 
The polarisation calibration process applied on our data is the standard LOFAR calibration adopted by the Magnetism Key Science Project (MKSP). For further details on the LoTSS polarised products, the readers may consult \cite{Shimwell2022, OSullivan2023} or past works, e.g. \cite{Vernstrom2018, OSullivan2020, Pomakov2022, OSullivan2023}.\\
During the calibration process, the integrated polarisation over the field of view is assumed to be negligible.
While this is usually a valid assumption at low frequencies, it does not hold completely in our field because of the presence of a strong polarised source. We nevertheless retained the standard assumption, with the caveat that it may lead to a bias in the polarisation properties of strong, compact sources. 
As we clarify below, this has no impact on our analysis, as we are interested in the polarised emission of the bridge.
The total frequency coverage ($120-168$ MHz) is subdivided into 480 channels, each of them $\sim 97$~kHz wide. Such high frequency sampling allows us to use the Rotation Measure \citep[RM;][]{B&B05} synthesis technique that minimizes bandwidth depolarisation in reconstructing the polarised emission \citep{B&B05}.
\subsection{Imaging}
\label{sec:imaging}
\begin{SCfigure*}[.2][t]
    \centering
    \includegraphics[scale=0.7]{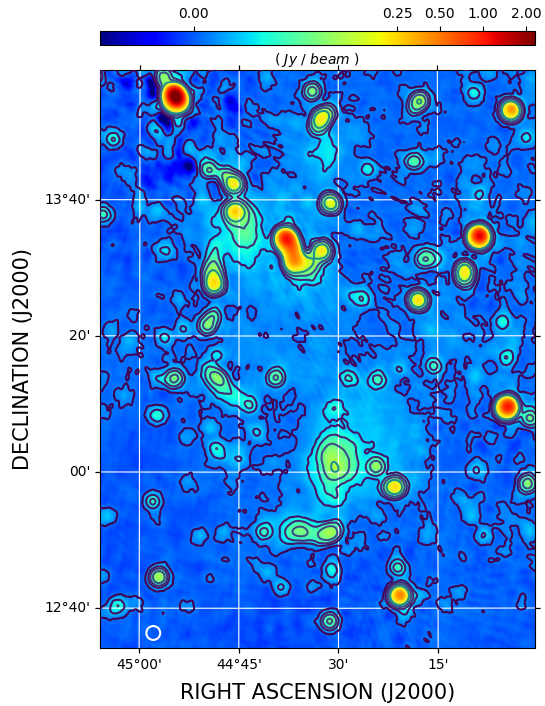}
    \caption{Image of the A399--A401 pair at 144~MHz. The $107'' \times 107''$ restoring beam is shown in white at the bottom left corner. Black contours are drawn at 4, 20, 40, 80, 160 and 320 $\times$ RMS, where the RMS is 620~$\mu \rm{Jy~beam^{-1}}$.}
    \label{fig:107asec_bridge}
\end{SCfigure*}
\begin{figure}
    \centering
    \includegraphics[scale=0.35]{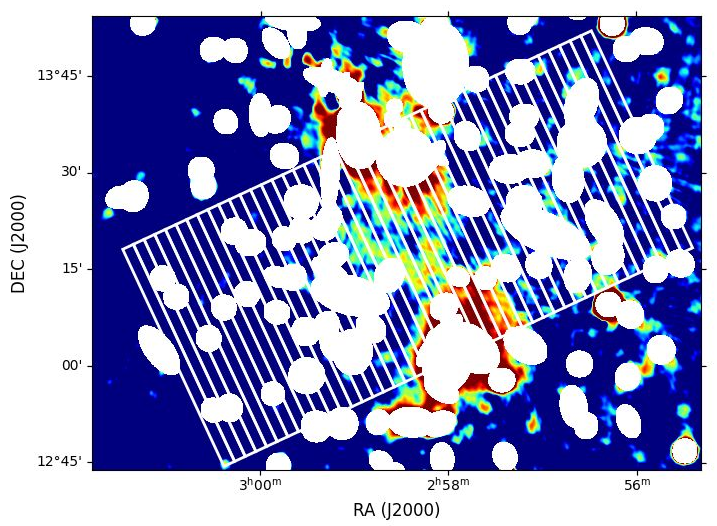}
    \includegraphics[scale=0.4]{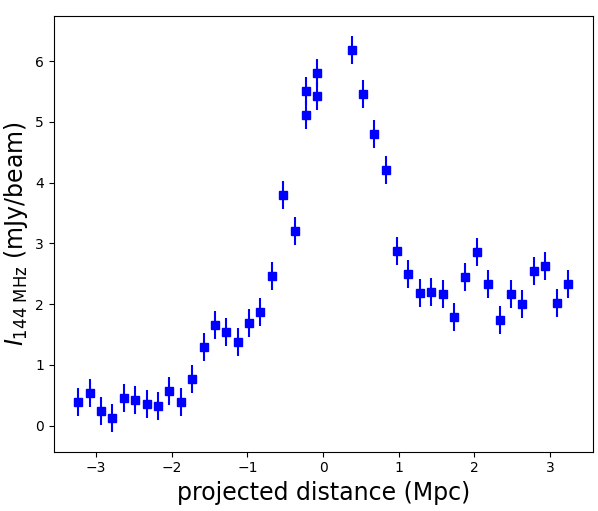}
    \caption{Bridge emission profile after masking of unrelated sources. Upper panel: Compact source subtracted image (see Sec.~\ref{sec:project} and Fig.~\ref{fig:107asec_bridge_img+region}) with non-bridge sources masked out (white ellipses). White line grid is used to measure the bridge brightness profile. The grid is tilted by 25$\degree$ and has the same reference point (centre) as \cite{Govoni19}. The lines are separated by 144~kpc (one beam width) and their length is $3$~Mpc. Lower panel: The surface brightness profile extracted by measuring the average surface brightness in each slice. 
    }
    \label{fig:bridge_profile}
\end{figure}

LoTSS pointings have a field of view of $3\degree - 4\degree$ (full width at half maximum), depending on the wavelength. The area covered by the bridge is $\sim 0.05 ~ \rm{deg^2}$, therefore, in order to speed up the subsequent steps, we select the data corresponding to the bridge region with a procedure called "extraction" \citep[see][]{vanWeeren2021}. 
We subtract the model components corresponding to sources outside the bridge as far out as 6~deg$^2$ from the visibilities.
Hence, our final dataset essentially contains the emission from the region close to the A399-A401 pair.
After the subtraction, we compute the primary beam correction towards the target coordinates and apply it directly to the visibilities. 
We then proceed to image both total intensity and polarised emission.
Calibrated visibilities are imaged using \textsc{WSClean v2.10.0}, which allows to perform multi-scale and multi-frequency deconvolution \citep{Wsclean14, Offringa2017}.
Initially, we make images at $6''$ and $20''$ angular resolution: 
as no evidence of bridge emission is visible at $20''$ resolution, we further reduce the resolution down to $107''$, where we obtain a clear detection of the bridge (Fig.~\ref{fig:107asec_bridge}).
Following \cite{Govoni19}, we extract the radial surface brightness profile as made by \cite{Govoni19} after subtracting sources embedded in the bridge region and masking out the sources not related to the radio ridge (white areas in Fig.~\ref{fig:bridge_profile}, detected using \texttt{PYBDSF} \citealt{pybdsf} when considering all the emission above 5~$\times$ RMS, an by visually inspecting the results to check whether there was residual unmasked emission clearly associated with compact source).
We report an average surface brightness of $ \langle I \rangle_{143 \rm{MHz}} = 4.29 \pm 0.06$~mJy~beam$^{-1}$ and a $ S_{143 \rm{MHz}} = 429 \pm 6$~mJy flux density for the bridge
(considering 100 beams in the selected region).
Compared with \cite{Govoni19}, our flux density and surface brightness measurements are $ \approx 20\%$ lower.
This difference may be due to the different calibration procedures and/or differences between the two observations, although it does not affect the remaining of our analysis.\\
The shortest baseline, ${\rm uv_{min}}$, of our observations is ${\sim 35}$~m (or ${\rm \sim 18 \lambda}$), implying that the largest angular size structure detectable is ${\rm \sim 3.5\degree}$. This is much larger than the projected distance between the two clusters (${\rm \sim 0.62\degree }$), therefore we do not expect that diffuse emission is filtered out.
We finally generate total intensity and polarisation images at $6''$, $20''$ and $107''$ respectively.
\subsection{Polarisation and RM synthesis recap}\label{Sec:pol_recap}

Synchrotron polarised radiation can be generally described in terms of Stokes $I$, $Q$, $U$ and $V$ parameters, which can be used to represent the orientation and intensity of the incoming electric field. We define the complex linear polarisation $P$ as:
\begin{equation}
    \label{Eq:Pol_vec}
    P = p I e^{2i\Psi} = Q + iU
\end{equation}
where $\Psi$ is the observed polarisation angle \citep[e.g.,][]{OSull2012}.
The degree of linear ($V=0$) polarisation $p$ is:
\begin{equation}
    p = \frac{\sqrt{Q^2 + U^2}}{I}
\end{equation}
and the polarisation angle $\Psi$ is:
\begin{equation}
    \Psi = \frac{1}{2} \arctan \frac{U}{Q}.
\end{equation}
When a linearly polarised wave propagates through a magnetised plasma extending over a path length $L$, the intrinsic polarisation angle $\Psi_0$ is rotated by an angle $\Delta \Psi$. 
This effect is called Faraday rotation and can be described by introducing the Faraday depth $\phi$ \citep{Burn1966,B&B05}:
\begin{equation}\label{Eq:phi-BB}
    \phi \simeq 0.81 \int_{\rm{source}}^{\rm{telescope}} n_e \, \boldsymbol{B} \cdot d\boldsymbol{l} \left[ \frac{\text{rad}}{\text{m}^2} \right],
\end{equation}
where $n_e$ is the electron number density in cm$^{-3}$, $\boldsymbol{B}$ the magnetic field in $\mu$G and $d\boldsymbol{l}$ is the infinitesimal path length in parsecs.
In the case where the polarised radiation is emitted by a background source and the rotation is only due to a foreground magneto-ionic medium, the source is defined Faraday thin and the variation in the polarisation angle $\Psi_{O}$ can be written as:
\begin{equation}\label{Eq:simple_psi_RM}
    \Psi_{obs}(\lambda) = \Psi_{0} + \phi \lambda^2.
\end{equation}
In this work, we adopt this technique for our analysis.

In the simplest case where only Faraday rotation occurs, the complex polarisation $P$ can be written as:
\begin{equation}\label{Eq:simple_P_RM}
    P(\lambda^2) = p_0 \, I e^{2i(\Psi_0 + \phi \lambda^2) },
\end{equation}
where $p_0$ is the intrinsic degree of polarisation of the synchrotron emission
and $\phi$ describes the Faraday rotation caused by the foreground magneto-ionic medium. 
The RM synthesis technique takes advantage of the similarity of Eq.~\ref{Eq:simple_P_RM} with a Fourier transform relationship by introducing the Faraday dispersion function (FDF) $F(\phi)$, or Faraday spectrum:
\begin{equation}
F(\phi) = K \int_{-\infty} ^{+\infty} P(\lambda^2) \, e^{-2i \phi (\lambda^2 - \lambda_0^2)} d\lambda^2, 
\end{equation}
where:
\begin{align}
    & K = \left( \int_{-\infty} ^{+\infty} W(\lambda^2) \, d\lambda^2 \right)^{-1} \\ 
    & \lambda_0^2 = \frac{\int_{-\infty} ^{+\infty} W(\lambda^2) ~ \lambda^2 d\lambda^2}{\int_{-\infty} ^{+\infty} W(\lambda^2) \, d\lambda^2} 
\end{align}
therefore we can measure/recover the amount of polarised emission of a source once its radiation has been de-rotated to a given Faraday depth.
\cite{B&B05} also introduce the Rotation Measure Transfer Function (RMTF), which is the Fourier transform of the wavelength sampling function ($W(\lambda^2)$).
In particular, a few specifications of an observation define some of the characteristics of the measured Faraday spectrum: 
the $\lambda^2$ coverage ($\Delta \lambda^2$) defines the resolution in Faraday space ($\delta \phi$), the wavelength resolution of the observation ($\delta \lambda^2$) sets the maximum observable Faraday depth ($\phi_{\rm{max}}$) and the largest scale in $\phi$ space to which one is sensitive depends on the shortest wavelength $(\lambda^2_{\rm{min}})$:
\begin{align}
    \delta \phi & \approx \frac{2 \sqrt{3}}{\Delta \lambda^2}\\
    ||\phi_{\rm{max}}|| & \approx \frac{\sqrt{3}}{\delta \lambda^2}
    \label{Eq:phi_max}\\
    \rm{max-scale} & \approx \frac{\pi}{\lambda^2_{\rm min}}
\end{align}
In our observations, the shortest  wavelength is $ \sim 1.79$~m, 
corresponding to a $\rm \sim 1.76~rad~m^{-2}$ max-scale\footnote{The max-scale set by our data would prevent us from detecting Faraday thick sources. 
However, this is not an issue since we will assume a Faraday thin bridge 
emission (Sec.~\ref{sec:depolarisation}). }, the resolution is $ \delta 
\phi \simeq 1.17$~rad~m$^{-2}$ and the maximum observable Faraday depth is 
$ || \phi_{\rm max} || \simeq 170$~rad~m$^{-2}$.\\
We compute the 
Faraday spectrum cube from the 107'' resolution polarisation images 
($\rm{ RMS_{Q,U} \sim 0.18 ~mJy~beam^{-1}}$) using a publicly available RM synthesis code 
\citep{RMsynth}\footnote{\texttt{https://github.com/CIRADA-Tools/RM-Tools}} and considering the average of the amplitude of the polarised emission over 
the bridge region (Fig.~\ref{fig:Fspec-bridge-107asec}).
\begin{figure}
    \centering
    \includegraphics[scale=0.3]{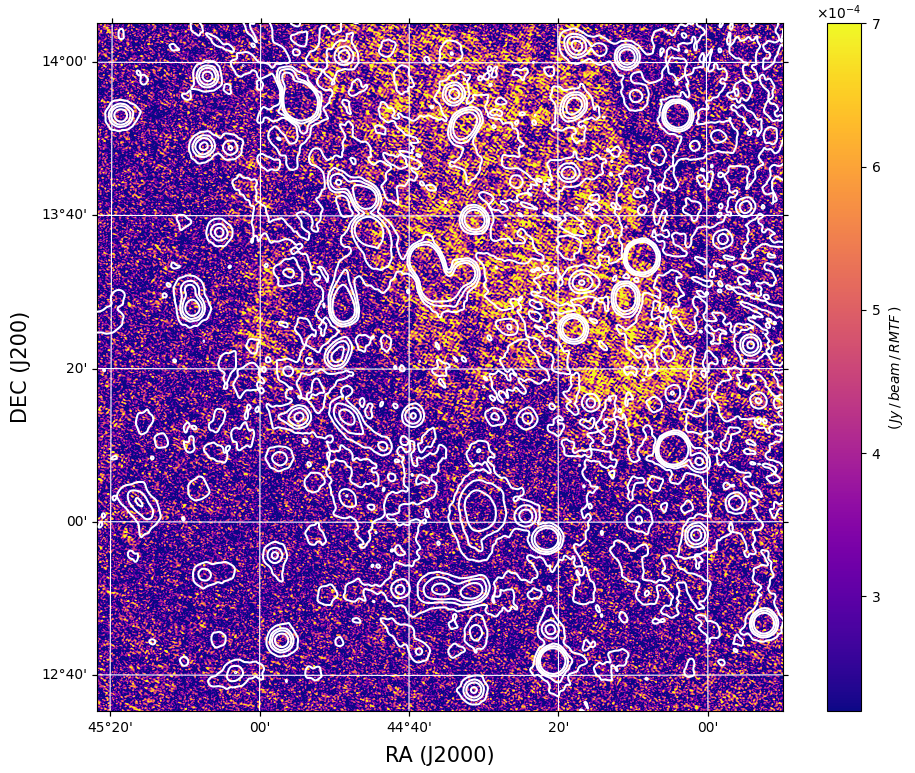}
    \caption{Slice of the FDF cube at $\phi \sim 7.5$~rad~m$^{-2}$ showing the Galactic polarised emission detected in the field. Total intensity is overlaid with contours at $4, 20, 40, 80 ~ \times {\rm RMS}$ taken from  Fig~\ref{fig:107asec_bridge}.}
    \label{fig:FDF_gal}
\end{figure}
\begin{figure}
    \centering
    \includegraphics[scale=0.18]{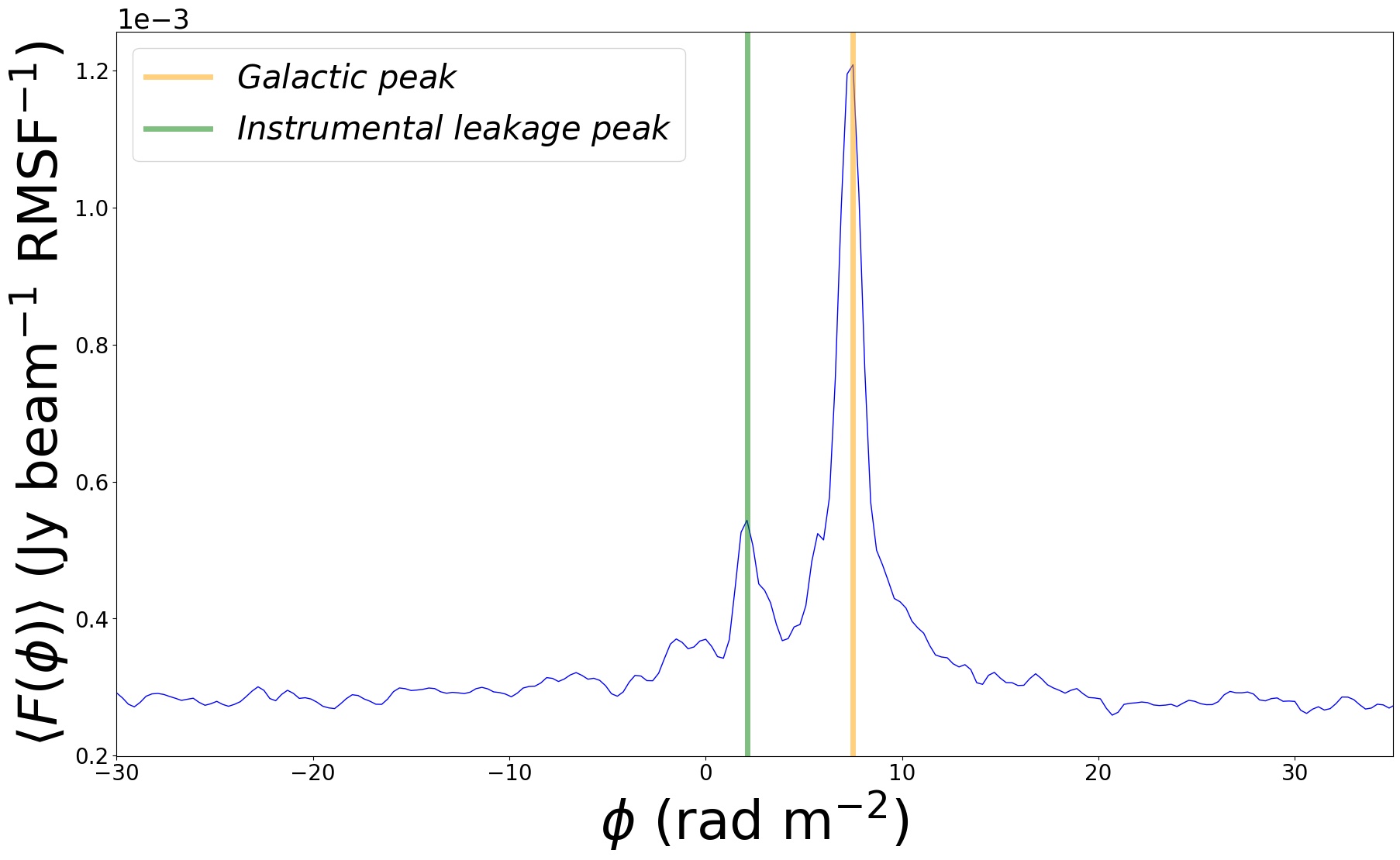}
    \caption{Faraday depth amplitude averaged over all the lines of sight within the bridge region at $107"$ resolution. It displays two peaks, which are due to the Galactic foreground medium and the instrumental leakage respectively.}
    \label{fig:Fspec-bridge-107asec}
\end{figure}
Most of the Faraday depth cube shows no emission, with the exception of diffuse emission that spans a large area of the field of view and peaks at $\phi \sim 7.5$~rad~m$^{-2}$ (Fig.~\ref{fig:FDF_gal}). This emission is largely uncorrelated with the total intensity emission and appears at relatively small Faraday depths, the typical characteristics of the Galactic emission at low frequencies \citep[e.g.,][]{Bernardi13,Van_Eck19,Erceg22}. We note that this Galactic foreground partially extends over the bridge region, although most of its emission appears outside of it. Fig.~\ref{fig:Fspec-bridge-107asec} shows the FDF amplitude averaged over the bridge region, defined as the 4$\sigma$ contour total intensity emission (Fig.~\ref{fig:107asec_bridge}).
The Galactic foreground peak is clearly visible, together with a fainter, second one at $ \phi= \phi_{\rm{instr}} \approx 2.1 ~\rm{rad~m^{-2}}$ likely due to instrumental leakage that is not corrected by our calibration procedure. Such instrumental feature is not uncommon in LOFAR observations and can appear up to $\pm 3$~rad~m$^{-2}$ \citep[e.g.,][]{OSullivan2019}.
No polarised emission from the bridge region is visible in the Faraday spectrum at any depth above the noise. We will use this lack of detection in the next sections to constrain the bridge emission mechanism and magnetic field.
\section{Polarisation analysis}
\label{sec:project}
The absence of polarised emission from the bridge can be used to place constraints on its magnetic field and, in turn, on its origin.
\subsection{Simulation expectations}
First, we test whether our observations are consistent with the prediction from the shock model proposed by \cite{Govoni19} to explain the origin of the bridge. The authors suggested that such inter-cluster radio emission may be the result of gas (re-)energisation and magnetic field amplification from several weak shocks ($M \sim 2-3$) originated in the initial merger phase. 
 In the adopted simulation of this system, a large distribution of weak shock waves, occupying a large volume filling factor ($\sim 1\%$) and an even larger surface filling factor ($\geq 30\%$) when seen in projection, forms as an effect of the supersonic turbulence developed in this region. These shocks may or may not emit synchrotron radiation, depending on the distribution of the relativistic electrons in the region. In particular, \cite{Govoni19} found that a pre-existing, mildly relativistic population of electrons could be re-accelerated by shocks and generate radio emission over the bridge extension that, in turn, ought to be polarised to some extent \citep[we refer the readers to][for further details about the emission model]{Govoni19}.
\begin{figure}
    \centering
    \includegraphics[scale=0.39]{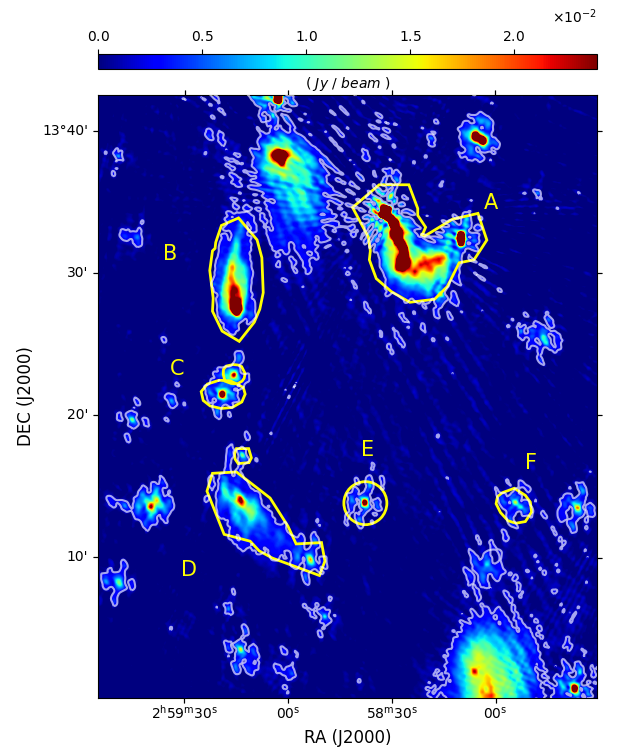}
    \caption{Zoom into the bridge field. Highlighted are the sources subtracted in order to retrieve the bridge emission model. Contours are drawn at $5 \times \rm{RMS}$, where RMS~$\sim 350$~$\mu$Jy~beam$^{-1}$. The image has a $20''$ resolution.}
    \label{fig:bridge+20asec_sources}
\end{figure}
\begin{figure}
    \centering
    \includegraphics[scale=0.33]{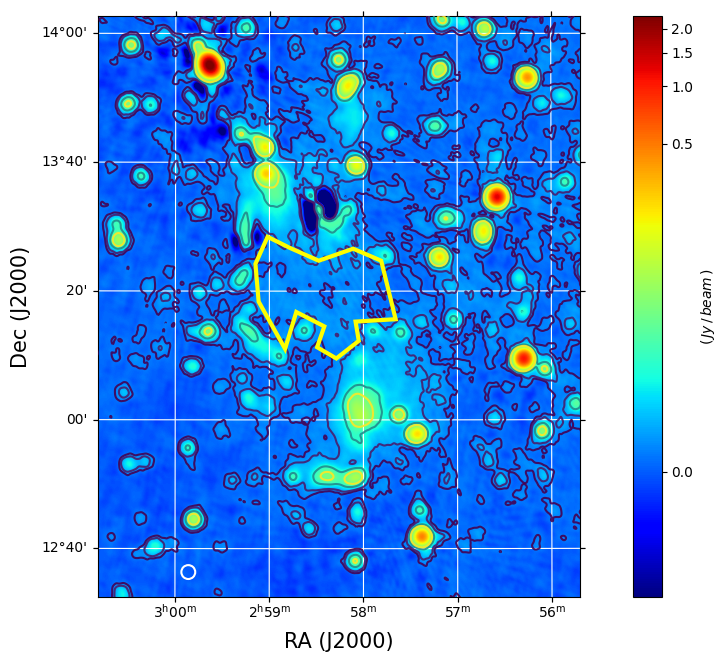}
    \caption{Image of the A399--A401 pair at 144~MHz once the sources within and around the bridge are subtracted (Fig.~\ref{fig:bridge+20asec_sources}). The region chosen to obtain the bridge emission model is highlighted in yellow. The $107'' \times 107''$ restoring beam is shown in white at the bottom left corner. The RMS of the image is $\sim 650$~$\mu$Jy~beam$^{-1}$, with contours at 4, 10, 40 and 80 $\times$ RMS. }
    \label{fig:107asec_bridge_img+region}
\end{figure}\\
A model of polarised emission from the bridge $P(x,y,\nu)$, where $(x,y)$ are sky directions and $\nu$ is the frequency, can be generated if the total intensity $I(x,y)$, the polarisation fraction $p(x,y)$ and the polarisation angle $\Psi(x,y)$ are known (Eq.~\ref{Eq:Pol_vec}).
We derive a total intensity template $I(x,y)$ from our total intensity observations. We first subtract bright sources located within or just around the bridge area (Fig.~\ref{fig:bridge+20asec_sources}), then we generate low-resolution, residual images where we eventually masked out all the pixels outside the bridge area defined by the 5$\sigma$ contours (Fig.~\ref{fig:107asec_bridge_img+region}).\\
Here, we compare our results with a cosmological simulation of a merging galaxy cluster. This simulation has been run with the ENZO-code an is part of a larger suite of cosmological simulations that have been extensively studied in \cite{Vazza2018} and \cite{Dominguez-Fernandez2019}. Here we analyse the cluster E5A, and for the details on the simulation, we point to the given references. The cluster E5A is undergoing an active merger, and throughout its formation, it hosts a bridge region, similar to the one in the A399-A401 system. Hence, it is an ideal candidate for comparison. 
Using a velocity-based shock finder \citep{Vazza09}, we can detect the shock waves in the simulation. Following the approach of \cite{WittorRR19}, we compute the polarised and unpolarised radio emission associated with electrons undergoing Diffusive Shock Acceleration in the bridge region. Eventually, we find a polarisation fraction close to the 70\% theoretical limit across the whole simulation volume \citep[see also][for details of the simulation]{Vazza2018,WittorRR19}. 
However, after convolving the Q and U simulated data cube to the resolution of our images, we obtain an average polarisation fraction of ${\sim 30 \%}$. Therefore, we assume a constant ${p = 30\%}$ across the bridge area (we also consider the limit case of an intrinsic polarisation of 10\%).\\
Finally, we assume, for simplicity, $\Psi(x,y) = 0$ and a constant Faraday depth $\phi$ across the bridge, i.e. the case that polarised emission is Faraday thin. In order to avoid any confusion with Galactic emission, we consider $\phi \gg \phi_{\rm gal} = 7.5$~rad~m$^{-2}$, specifically $\phi = 20$~rad~m$^{-2}$, constant across the bridge.
Under these assumptions, complex polarised emission $P$ can be generated following Eq.~\ref{Eq:simple_P_RM}: 
\begin{equation}
    P(x,y,\lambda) = p I(x,y,\lambda_0) \, e^{2 i \phi \lambda^2}.
\label{eq:complex_pol_rot_sim}    
\end{equation}
In practice, we want to simulate Stokes $Q$ and $U$ parameters as they are what we observe. 
We generate $Q$ and $U$ model images of the bridge following Eq.~\ref{eq:complex_pol_rot_sim}: 
\begin{align}
    Q (x,y,\lambda) = \Re{P(x,y,\lambda)} \nonumber \\
    U (x,y,\lambda) = \Im{P(x,y,\lambda)},
\end{align}
then Fourier transform and add (``inject'') them to our visibility data after point sources have been subtracted \citep[similar to the procedure in][]{Venturi08,Bonafede17,Locatelli2021,Nunhokee2023}. Visibilities are then imaged following the same procedure as for the real data, just averaging in frequency over 100 channels in order to reduce computing time. This choice reduces $\phi_{\rm max}$ to be $\sim 35$~rad~m$^{-2}$, still adequate for our simulations.\\
{Results of the injections are shown in the first panel of Fig.~\ref{fig:bridge_depo-FDFmap} and Fig.~\ref{fig:bridge_depo}. }The Faraday spectrum in Fig.~\ref{fig:bridge_depo} is different compared to Fig.~\ref{fig:Fspec-bridge-107asec} as we sum the amplitudes of the Faraday spectra over the bridge region:
\begin{equation}
    \hat{F}(\phi) =  \Sigma_{i}^N \left |F({\bf x}_i, \phi) \right |, 
\end{equation}
where ${\bf x}$ indicates the pixel coordinates and the sum runs over the total number of pixels $N$ of the bridge. From now on, we will refer to this quantity as the FDF.\\
{ Fig.~\ref{fig:bridge_depo} shows that the injected signal is detected in our data with a SNR~$\approx$ 100 and, if present in our data, it would have been clearly detected. Therefore, the absence of any polarised signal indicates that the emission generated in the weak shock scenario must be depolarised. Given the large scale of the emission, this is not surprising, however, under some assumptions, it allows us to put constraints on the magnetic field.}
\subsection{Constraints on the depolarisation mechanism}\label{sec:depolarisation}
\begin{figure*}
    \centering
    \includegraphics[scale=0.7]{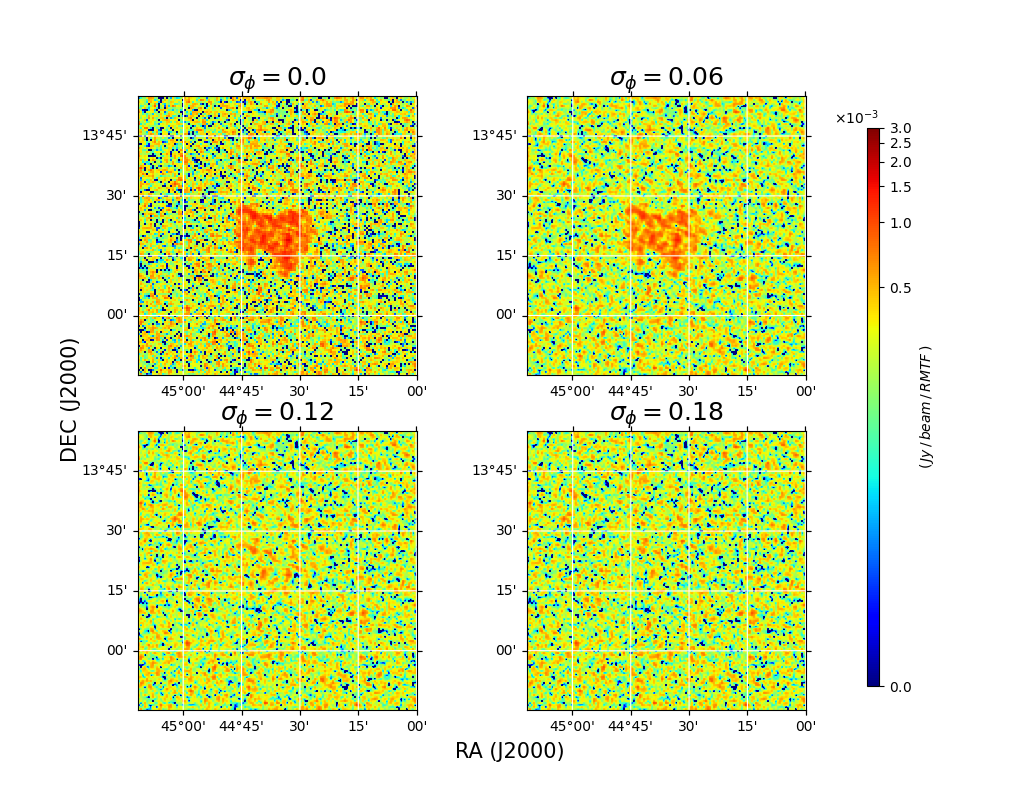}
    \caption{ Slices of the Faraday spectrum cube at { ${\phi = 20}$~rad~m$^{-2}$} as a function of Faraday dispersion $\sigma_{\phi}$, obtained from the injection of the bridge polarised emission model in the data - see text for details. It is evident that the bridge polarised emission decreases as the  Faraday dispersion increases until it, completely disappears when $ { \sigma_\phi = 0.18}$~rad~m$^{-2}$. The top left panel is the case where no depolarisation occurs, but only the rotation of the injected signal.
    }
    \label{fig:bridge_depo-FDFmap}
\end{figure*}
\begin{SCfigure*}[0.5][t!]
    \centering
    \includegraphics[scale=0.42]{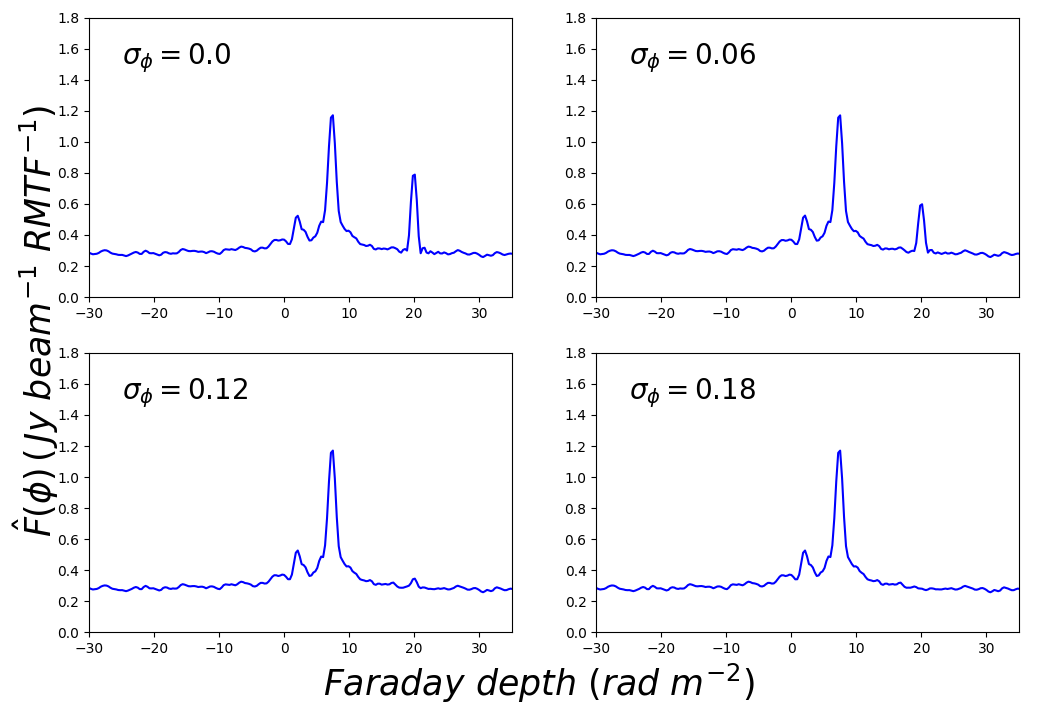}
    \caption[center]{Examples of FDFs of the bridge region as a function of Faraday dispersion $\sigma_\phi$, after injection of the depolarisation model in the data - see text for details. The peak at 20~rad~m$^{-2}$ decreases until it completely disappears below the nose for the $ { \sigma_\phi = 0.18}$~rad~m$^{-2}$ model.}
    \label{fig:bridge_depo}
\end{SCfigure*}
\begin{figure}
    \centering
    \includegraphics[scale=0.23]{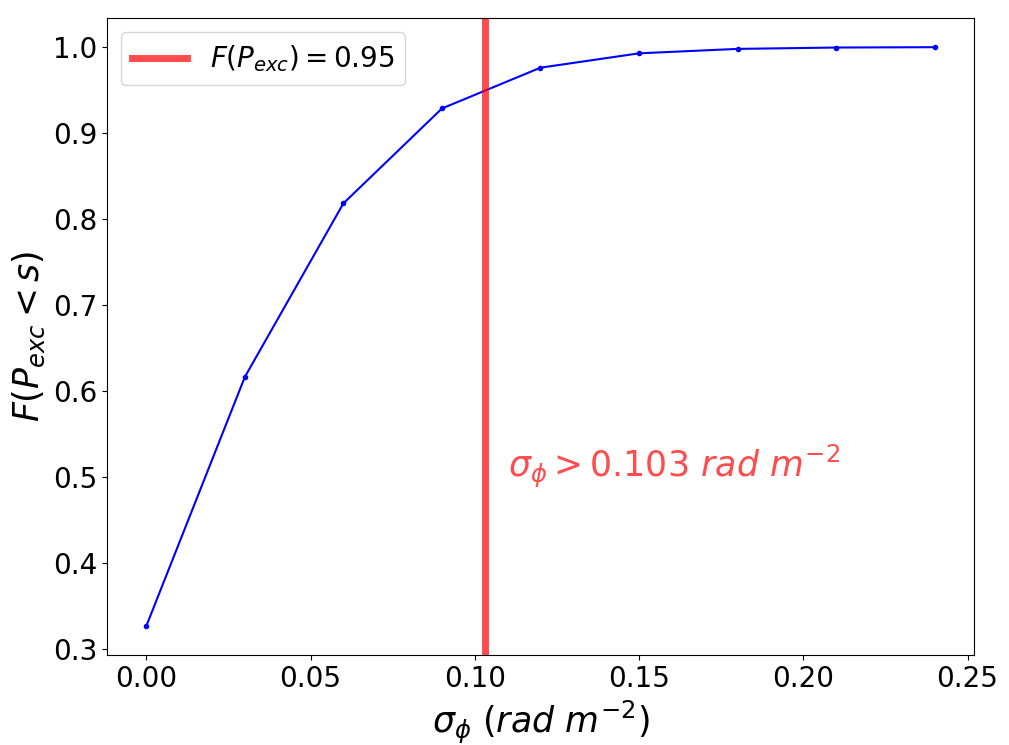}
    \caption{Normalised cumulative distribution ${F(P_{\rm{exc}})}$ as a function of ${\sigma_{\phi}}$, with the polarised emission from the bridge injected at ${\phi = 20 ~ \rm{rad ~ m^{-2}} }$. The red line indicates the threshold value of ${\sigma_{\phi}}$ at which the cumulative distribution function is 95\%.}
    \label{fig:cumulative-dist}
\end{figure}
Depolarisation refers to a process that reduces the intrinsic degree of polarisation of a source. Two typical cases of depolarisation are beam and bandwidth depolarisation. The first occurs when the polarisation angle changes significantly on scales smaller than the beam size. In this case, Stokes $Q$ and $U$ parameters change their sign and their integral over the beam area is smaller compared to the case when the polarisation angle is uniformly distributed.\\
Bandwidth depolarisation occurs when the polarisation angle changes significantly over the observing band - the case of high RM sources. In this case, similarly to the beam polarisation effect, the polarised intensity integrated over the bandwidth is smaller than the case where no rotation occurs.\\
The two aforementioned mechanisms are of instrumental origin, but depolarisation can have a physical origin and provide information of the physics of the source itself. A depolarisation effect that is of interest in our case occurs in the presence of a turbulent magnetic field in front of an emitting source, when spatial magnetic field fluctuations can be considered Gaussian. If the typical scale of turbulence of the magnetic field is smaller than the resolution element of the observation, the complex polarisation $P$ then becomes \citep[e.g.,][]{Sokoloff1998,OSull2012}:
\begin{equation}
    P = p \, I e^{-2 \sigma_{\phi}^2 \lambda^4} \, e^{2i(\Psi_0 + \phi \lambda^2) },
    \label{Eq:Fdepo-ext_Bturb}
\end{equation}
where $\sigma_\phi$ is the Faraday dispersion, which quantifies spatial fluctuations of the Faraday depth due to the magnetic field variations. We note that the polarisation amplitude is reduced by a factor $e^{-2 \sigma_{\phi}^2 \lambda^4}$ with respect to Eq.~\ref{eq:complex_pol_rot_sim}, which is strongly wavelength dependent. This case is referred to as depolarisation due to external Faraday dispersion \citep[e.g.][]{Tribble1991}.
We consider external Faraday dispersion as a depolarisation mechanism in our case. In particular, we retain the assumption that the polarised emission is Faraday thin, i.e. the shock width is much smaller than the bridge. In addition, our observations are not sensible to Faraday thick structures, therefore we dismiss the case of internal Faraday depolarisation. In this framework, the magnetic field fluctuations in front of the shocks are responsible for the depolarisation.
Our observations can, therefore, place a lower limit on $\sigma_\phi$, i.e. on the minimum magnetic field fluctuation in the foreground screen necessary to completely depolarise the bridge signal.\\
We follow the same procedure described earlier to generate Stokes $Q$ and $U$ parameters, with the only difference that we added the depolarisation term to Eq.~\ref{eq:complex_pol_rot_sim}, i.e.:
\begin{equation}
    P(x,y,\lambda) = p I(x,y,\lambda_0) \, e^{-2 \sigma_{\phi}^2 \lambda^4} \, e^{2 i \phi(x,y) \lambda^2},
\label{eq:complex_pol_rot_sim_depol}    
\end{equation}
where, in this case, $\phi(x,y)$ is a realisation drawn from a Gaussian distribution with a 20~rad~m$^{-2}$ mean and a standard deviation $\sigma_\phi$. 
Performing the same injection steps previously described, we obtain a Faraday spectrum cube for each value of ${\sigma_{\phi}}$ taken in the $ 0 - 0.24$~rad~m$^{-2}$ range, with steps of 0.03~rad~m$^{-2}$.
The results for a few, selected ${\sigma_\phi}$ values are shown in Fig.~\ref{fig:bridge_depo-FDFmap} and Fig.~\ref{fig:bridge_depo}.
The $\sigma_{\phi} = 0$~rad~m$^{-2}$ case has been discussed previously and shows that the polarised emission from a weak shock model should be visible in our data in case of Faraday rotation without depolarisation. As ${\sigma_{\phi}}$ increases, the polarised emission quickly decreases, until it completely disappears when ${\sigma_{\phi} = 0.18}$~rad~m$^{-2}$. This is also evident in FDF profile, which becomes consistent with noise in the ${\sigma_{\phi} = 0.18}$~rad~m$^{-2}$ case. 
These results imply that a minimum value ${\sigma_{\phi} = \sigma^* < 0.18}$~rad~m$^{-2}$ must, therefore, exist below which the signal is not sufficiently depolarised and should be visible in our observations.
In order to estimate such a lower limit on ${\sigma_{\phi}}$, we follow a procedure similar to \cite{Nunhokee2023} and calculate the cumulative distribution function of the following ratio:
\begin{equation}\label{Eq:P_exc}
    P_{\rm{exc}} (\sigma_\phi) = \frac{\int_{\phi_1}^{\phi_2} \hat{F}_{\rm{inj}} (\sigma_\phi,\phi) \, d\phi - \int_{\phi_1}^{\phi_2} \hat{F}_{\rm{o}}(\phi) \, d\phi }{\int_{\phi_1}^{\phi_2} \hat{F}_{\rm{o}}(\phi) \, d\phi},
\end{equation}
where ${\hat{F}_{\rm{inj}}}$ and ${\hat{F}_{\rm{o}}}$ are the injected and observed FDF, respectively, and ${\phi_1 = 15}$~rad~m$^{-2}$ and ${\phi_1 = 25}$~rad~m$^{-2}$. In other words, the ratio ${P_{\rm{exc}}}$ is the excess of the injected polarised emission with respect to the data as a function of ${\sigma_\phi}$, calculated in a region centred on the average injected Faraday depth $ {\phi = 20}$~rad~m$^{-2}$.

We note that: 
\begin{align}
        &\lim_{\sigma_{\phi}\to0} P_{\rm{exc}} = P' < \infty\\
        & \lim_{\sigma_{\phi}\to\infty} P_{\rm{exc}} = 0,
\end{align}
i.e., $P_{\rm{exc}}$ is a monotonically decreasing function of $\sigma_{\phi}$. 
We find a probability ${ F(P_{\rm{exc}} < 0.95)}$ for ${\sigma_{\phi}^* = 0.10}$~rad~m$^{-2}$ (Fig.~\ref{fig:cumulative-dist}), in other words, if the ${ \sigma_{\phi}} $ was smaller than 0.10~rad~m$^{-2}$, polarised emission should be detected in our data with a 95\% confidence level (or greater). As there is no detection, our observations set a limit on the Faraday dispersion ${ \sigma_{\phi} > 0.10}$~rad~m$^{-2}$ at 95\% confidence level.
In the next section, we turn this lower limit into a lower limit on the magnetic field in the bridge.
\section{Constraints on the bridge magnetic field}
\label{sec:constraints_mag_field}
\begin{figure}
    \centering
    \includegraphics[scale=0.345]{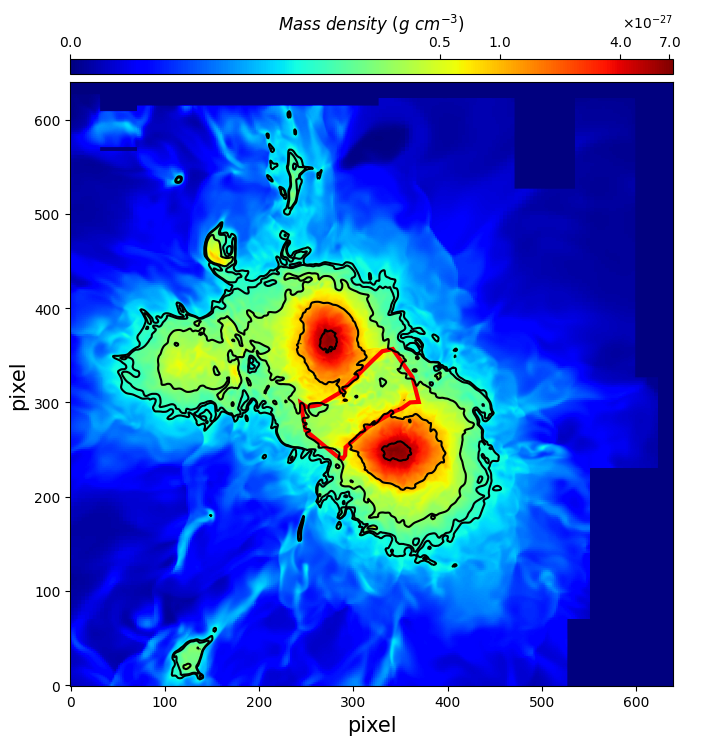}
    \caption{Central slice of the cube density from simulations. Moving towards the density cube, the average density of the medium initially increases, indicating the presence of the cluster pair, then starts to decrease until reaching $0$ in the last slice, showing the ending regions of the cluster pair. As so, we selected the central slice of the cube as the best indicator of the density distribution. The density contours are at $1, \, 1.24, \, 2.61, \, 10.3, \, 53.5 \times 10^{-28}$ g~cm$^{-3}$. In red, is also highlighted the region selected as the bridge region in the simulation. Every pixel is $\sim 16$ kpc long.}
    \label{fig:density_ctr+bridge_reg-sim}
\end{figure}
\begin{figure}
    \centering
    \includegraphics[scale=0.3]{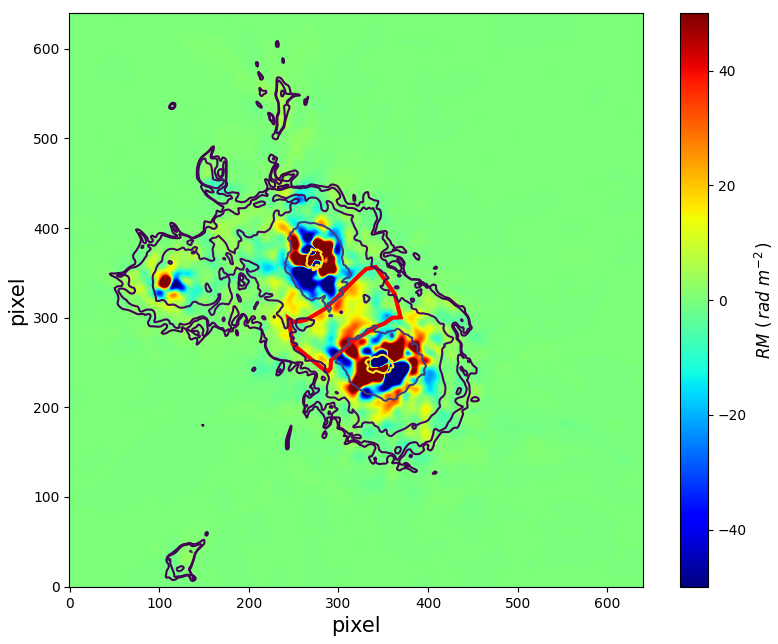}
    \caption{Faraday depth map obtained using the cluster simulation \citep{Wittor2019} - see text fr details. The contours are from the density field in Fig.~\ref{fig:density_ctr+bridge_reg-sim}, and the selected bridge region is highlighted in red.  Every pixel is $\sim 16$ kpc long.}
    \label{fig:RM_map_sim}
\end{figure}
\begin{figure}[h!]
    \centering
    \includegraphics[scale=0.32]{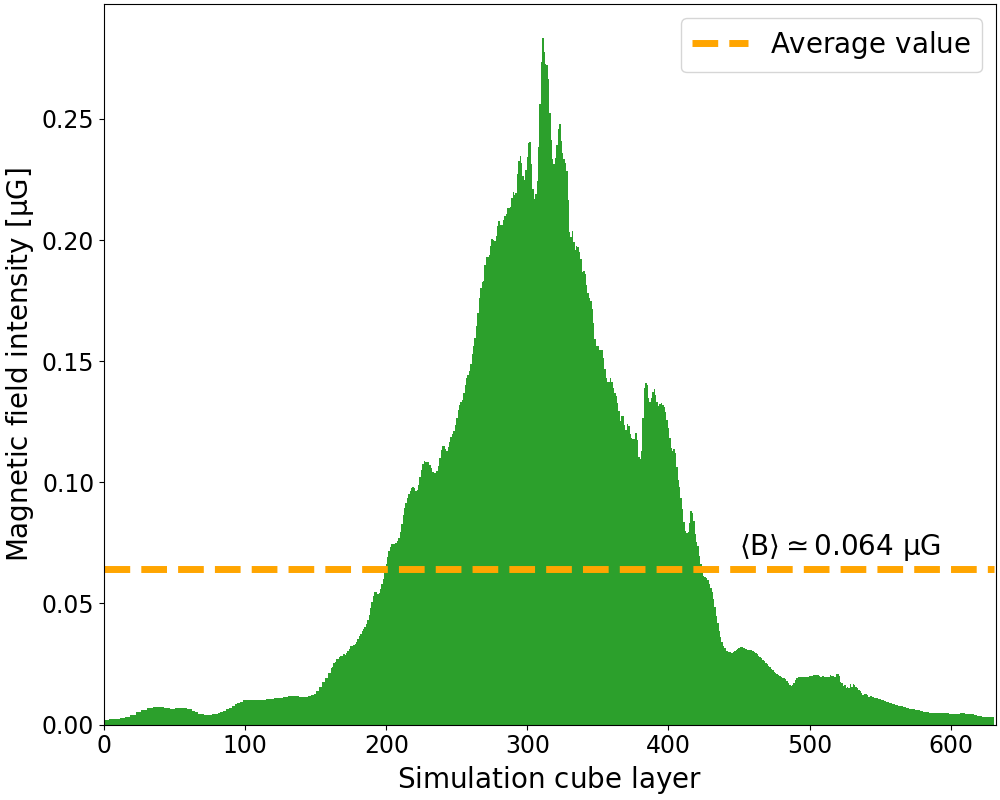}
    \caption{Distribution of the density-weighted average magnetic field intensity of each layer of the simulation cube. The average value of all the layers is indicated by the orange line. 
    }
    \label{fig:B_hist}
\end{figure}
The lower limit on the Faraday dispersion $\sigma_\phi$ can be translated into a limit on the magnetic field using the relation between Faraday depth and magnetic field (Eq.~\ref{Eq:phi-BB}). The standard deviation $\sigma_\phi$ of the spatial distribution of the Faraday depth $\phi$ is defined as:
\begin{equation}\label{Eq:sigma_phi_B}
    \sigma_\phi = \sqrt{\langle \phi^2({\bf x}) \rangle - \langle \phi({\bf x}) \rangle^2}
 \end{equation}
where $\bf x$ indicates the line of sight (or spatial coordinate) and $\langle \rangle$ is the average ensemble. By using the definition of Faraday depth (Eq.~\ref{Eq:phi-BB}), Eq.~\ref{Eq:sigma_phi_B} becomes:
\newpage
\begin{equation}\label{Eq:sigma_phi_B1}
    \sigma_\phi = 0.81 \left [\left \langle \left ( \int_L n_e({\bf x}) \, \boldsymbol{B}({\bf x}) \cdot d {\bf l} \right )^2 \right \rangle - \left( \left \langle \int_L n_e({\bf x}) \, \boldsymbol{B}({\bf x}) \cdot d {\bf l} \right \rangle \right)^2  \right ] ^{1/2}
\end{equation}
where $L$ becomes, in our case, the line-of-sight depth of the bridge. Eq.~\ref{Eq:sigma_phi_B1} shows that the standard deviation of the Faraday depth is a function of the (density-weighted) fluctuations of the magnetic field, therefore, a lower limit on the standard deviation of the Faraday depth imposes a lower limit on the spatial fluctuations of the magnetic field. In order to derive such a lower limit we make a few simplifying assumptions. We consider the electron density not to be a free parameter, but we assume it from the cluster simulation \citep{Dominguez-Fernandez2019,Wittor2019}. We also assume the magnetic field from the cluster simulation but we allow it to be scaled by an overall, spatially independent factor $\alpha$. A theoretical Faraday dispersion $\sigma_{\phi,m}$ can therefore be computed from the simulation:   
\begin{eqnarray}
\label{Eq:sigma_phi_B2}
    \sigma_{\phi,m} & = & 0.81 \left[ \left \langle \left ( \int_L n_{e,s}({\bf x}) \, \alpha \boldsymbol{B}_s ({\bf x}) \cdot d {\bf l} \right )^2 \right \rangle  - \right. \nonumber \\
    & &\left. \left( \left \langle \int_L n_{e,s} ({\bf x}) \, \alpha \boldsymbol{B}_s ({\bf x}) \cdot d {\bf l} \right \rangle \right)^2  \right]^{1/2} \nonumber \\
    & = & \alpha \left ( 0.81 \left[\left \langle \left ( \int_L n_{e,s}({\bf x}) \, \alpha \boldsymbol{B}_s ({\bf x}) \cdot d {\bf l} \right )^2 \right \rangle - \right. \right. \nonumber \\
    & & \left. \left. \left( \left \langle \int_L n_{e,s} ({\bf x}) \, \alpha \boldsymbol{B}_s ({\bf x}) \cdot d {\bf l} \right \rangle \right)^2 \right]^{1/2} \right ) \nonumber \\
    & = & \alpha \, \sigma_{\phi,s},
\end{eqnarray}
where the subscript $s$ indicates all quantities derived from the cluster simulation and $\alpha$ is the free parameter constrained by the observational lower limit on $\sigma_\phi$, i.e.:
\begin{equation}
\label{Eq:sigma_phi_B3}
    \sigma_{\phi,m} = \alpha \, \sigma_{\phi,s} \geq \sigma_\phi^*.
\end{equation}
The standard deviation of the Faraday depth from the cluster simulation is computed by identifying a region similar to the observed bridge.
Here, we consider the electron density cube of the \cite{Wittor2019} simulation (shown by white contours in Fig.~2 in their work). We then select the inter-cluster region so that it has a physical dimension equal to the one defined through observations. The final region is highlighted in red in Fig.~\ref{fig:density_ctr+bridge_reg-sim}).
We compute the Faraday depth for each pixel of the selected region by integrating the density-weighted magnetic field and then we smooth the Faraday depth map at the same resolution as the observed bridge, i.e. $107''$, corresponding to a 144~kpc physical size (Fig.~\ref{fig:RM_map_sim}). 
We derive the simulated Faraday dispersion ${ \sigma_{\phi,s}} $ from the map and substitute it into Eq.~\ref{Eq:sigma_phi_B3} in order to obtain $ { \alpha \ge 7.2 \times 10^{-3}}$ (where as $\sigma_{\phi, m}$ we use the observational limit derived in Sec.~\ref{sec:depolarisation}). In other words, the simulated Faraday dispersion is already higher than the observed lower limit, consistent with the injection results from Sec.~\ref{sec:project}.\\
The lower limit on the model parameter $\alpha$ can be turned into a lower limit on the average magnetic field in the bridge region, indeed assuming that $\alpha$ is a scaling factor of the magnetic field in the cluster simulations (Eq.~\ref{Eq:sigma_phi_B2}). 
In order to find the average magnetic field along the bridge extension, we first compute the total magnetic field intensity in each pixel from the simulated cube.
We then derive the density-weighted, average magnetic field intensity for each layer of the cube and, eventually, averaged these values across all layers, obtaining a mean magnetic field intensity of $\langle B_{\rm b,s} \rangle = 0.064$~$\mu$G for the bridge region.
The distribution of the magnetic field intensity along the selected bridge region in the simulation cube is shown in Fig.~\ref{fig:B_hist}.
This procedure allows us to set a lower limit on the mean magnetic field in the bridge region $\hat{B}_m$:
\begin{equation}
 {  \hat{B}_m \ge \alpha \, \langle B_{\rm b,s} \rangle \ge 0.46 \, {\bf \rm nG}.
}
\end{equation}
We also derive the magnetic field limit by considering a lower limit polarised fraction of 10\%. This turns into a lower limit  ${\sigma_{\phi} > 0.09}$ and a consequent magnetic field limit ${B_{\rm b,s} \ge 0.41}$ nG.

\subsection{Instrumental and Galactic peak injection}\label{sec:inst-gal_inj}

Until now, we have injected the bridge at $\phi = 20$~rad~m$^{-2}$, away from the instrumental and Galactic Faraday depth peaks. In this section, we relax this assumption in order to derive upper limits on the Faraday dispersion and bridge magnetic field in the presence of Galactic and instrumental ``contamination''.\\
We follow the same injection procedure described in Sec.~\ref{sec:depolarisation} and \ref{sec:constraints_mag_field}, simulating Stokes $Q$ and $U$ visibilities from the cluster simulations, only this time at $\phi = 7.5$~rad~m$^{-2}$, the Faraday depth where the maximum of the Galactic emission appears. 
By applying the same procedure as in the previous sections, we are assuming in the real Faraday spectrum (Fig.~\ref{fig:Fspec-bridge-107asec}) no significant contribution from the bridge and that the polarised emission is due only to the Galactic foreground. 
As shown in Fig.~\ref{fig:FDF_gal}, the galaxy polarised emission is clearly prevalent in the field. Furthermore, from this image, it is evident that when we increase the region within which the Faraday spectrum is being extracted, the Galactic polarised peak becomes even more dominant than in Fig.~\ref{fig:Fspec-bridge-107asec}. 
In conclusion, we cannot completely rule out a minimal contribution from bridge polarised emission, but it is reasonable to assume that such emission is negligible with respect to the strong Galactic one.\\
As in the previous analysis, we constructed a cumulative probability function for depolarisation models that samples the ${ 0 \le \sigma_\phi \le 0.24}$~rad~m$^{-2}$ range (same as Sec.~\ref{sec:depolarisation}) and set an upper limit to the Faraday dispersion ${\sigma_\phi \ge 0.09}$~rad~m$^{-2}$ at 95\% confidence level.\\ 
\\
We repeat the same procedure for the instrumental case, where we injected the simulated polarised signal at ${ \phi = 2.1}$~rad~m$^{-2}$, the Faraday depth at which the instrumental leakage appears. In this case, we set an upper limit ${ \sigma_\phi \ge 0.10}$~rad~m$^{-2} $ at 95\% confidence level.\\
\\
Both limits are somewhat lower than the case where the signal was injected in the featureless part of the Faraday spectrum, as qualitatively expected. If we take the lowest between the two limits, we derive a slightly lower limit on the mean magnetic field of the bridge, ${ \hat{B} \ge 0.41}$~nG.\\
\\
It is useful to compare our upper limits to the standard estimate that can be obtained via the standard equipartition assumption, i.e. of a minimum energy of the relativistic plasma. The equipartition magnetic field $B_{\rm eq}$ can be derived following \citep[][Eq.~25 and 26]{Govoni04}, and assuming a spectral index $\alpha=1.4$, a bridge size of 1~Mpc and a unity filling factor. We obtained 
$B_{eq} = 0.24$~$\mu$G, well above our lower limit. We also note that,
considering the thin shock scenario, the filling factor is likely to be smaller than one, leading to a higher equipartition magnetic field, still consistent with our limit. 
Finally, it is also worth noting that the equipartition magnetic field is approximately three times larger than values found in inter-cluster regions \citep{Hoang2023}.
\section{Discussion and conclusions}\label{sec:conclusions}
\begin{figure}
    \centering
    \includegraphics[scale=0.5]{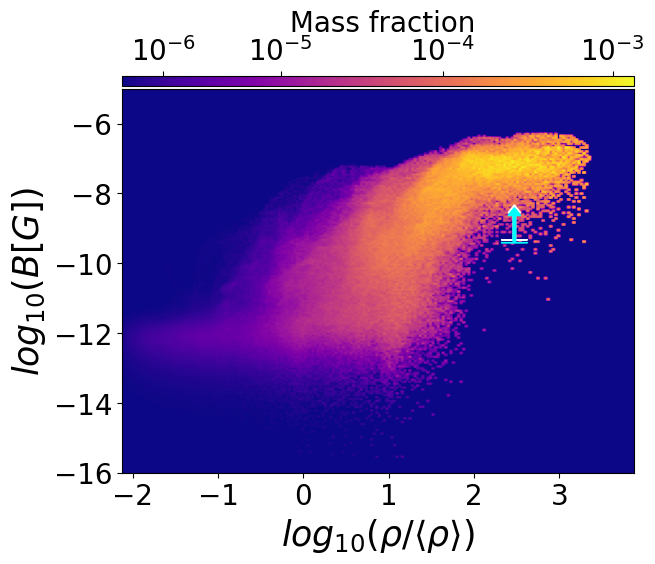}
    \caption{ Phase diagram of the statistical distribution of the magnetic field strength as a function of gas overdensity ($|\boldsymbol{B}| - \rho / \langle \rho \rangle$) from simulations \citep{Vazza2021}. 
    The cyan and the white arrows are the ${0.41 ~ {\rm and } ~ 0.46}$~nG lower limits on the magnetic field from this work.}
    \label{fig:magnetic_predictions}
\end{figure}
In this paper, we have presented a radio polarisation study of the merging cluster pair Abell~399--Abell~401. \cite{Govoni19} has studied this system through LOFAR observations at 144~MHz, showing the first evidence of a radio bridge connecting the two clusters. 
The authors suggested that a shock-driven emission model, where multiple weak shocks, originated by the motion of the clusters during the merging event, re-accelerates a pre-existing population of electrons, triggering the radio emission. Considering the proposed scenario, numerical simulations suggested that the emission can be $ {\sim 70\%} $ polarised \citep{Wittor2019}. However, at the resolution of our observations, the observed polarised emission is expected to be lower, ${\sim 30\%}$.
We have imaged LOFAR observations of the bridge region at 144~MHz at 
$6''$, $20''$ and $107''$ angular resolution respectively. Total intensity images at $6''$ and $20''$ showed only emission from compact sources 
whereas the bridge emission is clearly evident in images with $107''$ angular resolution, essentially confirming results from \cite{Govoni19} and, more recently, \cite{deJong2022}.
In order to search for polarised emission from the bridge, we 
performed the RM synthesis analysis on the $107''$ resolution images. We did not reveal any significant polarised emission from the bridge,
but only from Galactic foreground and instrumental leakage.
We therefore used our observations to set an upper limit on the bridge polarised emission.\\
We assumed the model used by \cite{Govoni19} to justify the bridge emission and the simulation of \cite{Wittor2019} to compute the expected polarised fraction. Accounting also for the beam geometric depolarisation, we found that the polarised emission expected from the simulation should have been easily detectable in our observations, suggesting the presence of a depolarisation mechanism.
Under the assumption that the shock width is negligible with respect to the bridge extension, depolarisation is due to the remaining portion of the bridge that acts as an external Faraday screen.
We derived a lower limit on the dispersion of the external Faraday screen ${\sigma_{\phi} \ge 0.10}$~rad~m$^{-2}$ (${ \ge 0.09}$~rad~m$^{-2}$, if the bridge polarised emission falls near the FDF Galactic peak), which, in turn, becomes a lower limit on the mean magnetic field of the bridge ${\hat{B}_m \ge 0.46}$~nG ($ {\ge 0.41}$~nG).
We stress that this lower limit is valid for the bridge medium that acts as an external Faraday screen. This is not a limit on the radio-emitting regions (shock regions). However, for such emitting regions, we would expect: (i) a higher (0.2-0.4~$\rm{\mu G}$) magnetic field strength which, if taken into account would only increase the average ${B}$ value; (ii) shock emitting regions, in our approximation, are small regions, so our lower limit is still representative for the majority of the inter-cluster medium. Therefore, in the framework of an external Faraday screen originated by the bridge, we are, again, in a conservative condition.\\
Assuming a bridge mean density\footnote{We notice that the recent study of \cite{Hincks2021-bis} on the A399-A401 cluster pair through the SZ effect, reported a much lower density estimate for the bridge region ($(0.88 \pm 0.24) \times 10^{-4}$~cm$^{-3}$). Such a low density value is the result of a larger estimate of the bridge length, which, following the authors, is observed at a small angle with respect to the line of sight. Despite our technique could, in principle, be applied also to such a scenario, we could not include it in our study because the used simulation refers to an epoch at which the merging clusters are separated by $\sim 3.5$~Mpc.} of $\sim 3.4 \times 10^{-4}$~cm$^{-3}$ \citep{Fujita2008}, the corresponding overdensity is $\delta \sim 300$.
Fig.~\ref{fig:magnetic_predictions} shows predictions for the magnetic field intensity as a function of the gas overdensity \citep{Vazza2021}. Almost all models predict a magnetic field intensity greater $\sim 1$~nG at an overdensity of $\sim 300$, largely consistent with our constraints. 
\section*{Acknowledgements}
We thank the referee Steve Spangler for the valuable comments and suggestions on our work.\\
LOFAR \citep{LOFAR2013} is the Low Frequency Array designed and constructed by ASTRON. It has observing, data processing, and data storage facilities in several countries, which are owned by various parties (each with their own funding sources), and thatare collectively operated by the ILT foundation under a joint scientific policy. The ILT resources have benefited from the following recent major funding sources: CNRS-INSU, Observatoire de Paris and Université d'Orléans, France; BMBF, MIWF-NRW, MPG, Germany; Science Foundation Ireland (SFI), Department of Business, Enterprise and Innovation (DBEI), Ireland; NWO, The Netherlands; The Science and Technology Facilities Council, UK; Ministry of Science and Higher Education, Poland; The Istituto Nazionale di Astrofisica (INAF), Italy. This research made use of the Dutch national e-infrastructure with support of the SURF Cooperative (e-infra 180169) and the LOFAR e-infra group. The Jülich LOFAR Long Term Archive and the GermanLOFAR network are both coordinated and operated by the Jülich Supercomputing Centre (JSC), and computing resources on the supercomputer JUWELS at JSC were provided by the Gauss Centre for Supercomputinge.V. (grant CHTB00) through the John von Neumann Institute for Computing (NIC). This research made use of the University of Hertfordshirehigh-performance computing facility and the LOFAR-UK computing facility located at the University of Hertfordshire and supported by STFC [ST/P000096/1], and of the Italian LOFAR IT computing infrastructure supported and operated by INAF, and by the Physics Department of Turin university (under anagreement with Consorzio Interuniversitario per la Fisica Spaziale) at the C3S Supercomputing Centre, Italy.\\
Annalisa Bonafede acknowledges support from the ERC-Stg DRANOEL n.714245 and from the MIUR grant FARE "SMS". D.W. is funded by the Deutsche Forschungsgemeinschaft (DFG, German Research Foundation) - 441694982. The authors gratefully acknowledge the Gauss Centre for Supercomputing e.V. (www.gauss-centre.eu) for supporting this project by providing computing time through the John von Neumann Institute for Computing (NIC) on the GCS Supercomputer JUWELS at Jülich Supercomputing Centre (JSC), under project no. hhh44. F.V.  acknowledges financial support from the Horizon 2020 program under the ERC Starting Grant MAGCOW, no. 714196. Our simulations were run on the Piz Daint supercomputer at CSCS-ETH (Lugano), on the  JUWELS cluster at Juelich Superc omputing Centre (JSC), under projects “radgalicm"  and on   the Marconi100 clusters at CINECA (Bologna), under project INA21. ABotteon acknowledges support from the ERC-StG DRANOEL n. 714245. RJvW acknowledge support from the VIDI research programme with project number 639.042.729, which is financed by the Netherlands Organisation for Scientific Research (NWO). V.V. acknowledges support from INAF mainstream project “Galaxy Clusters Science with LOFAR” 1.05.01.86.05. \\ \\
Data used in this work are available upon reasonable request to the authors.


%
\bibliographystyle{aa} 
\bibliography{biblio.bib} 

\begin{thebibliography}{73}
\expandafter\ifx\csname natexlab\endcsname\relax\def\natexlab#1{#1}\fi

\bibitem[{{Akamatsu} {et~al.}(2017){Akamatsu}, {Fujita}, {Akahori}, {Ishisaki},
  {Hayashida}, {Hoshino}, {Mernier}, {Yoshikawa}, {Sato}, \&
  {Kaastra}}]{Akamasu2017}
{Akamatsu}, H., {Fujita}, Y., {Akahori}, T., {et~al.} 2017, \aap, 606, A1

\bibitem[{{Bernardi} {et~al.}(2013){Bernardi}, {Greenhill}, {Mitchell}, {Ord},
  {Hazelton}, {Gaensler}, {de Oliveira-Costa}, {Morales}, {Udaya Shankar},
  {Subrahmanyan}, {Wayth}, {Lenc}, {Williams}, {Arcus}, {Arora}, {Barnes},
  {Bowman}, {Briggs}, {Bunton}, {Cappallo}, {Corey}, {Deshpande}, {deSouza},
  {Emrich}, {Goeke}, {Herne}, {Hewitt}, {Johnston-Hollitt}, {Kaplan}, {Kasper},
  {Kincaid}, {Koenig}, {Kratzenberg}, {Lonsdale}, {Lynch}, {McWhirter},
  {Morgan}, {Oberoi}, {Pathikulangara}, {Prabu}, {Remillard}, {Rogers},
  {Roshi}, {Salah}, {Sault}, {Srivani}, {Stevens}, {Tingay}, {Waterson},
  {Webster}, {Whitney}, {Williams}, \& {Wyithe}}]{Bernardi13}
{Bernardi}, G., {Greenhill}, L.~J., {Mitchell}, D.~A., {et~al.} 2013, \apj,
  771, 105

\bibitem[{{Bonafede} {et~al.}(2017){Bonafede}, {Cassano}, {Br{\"u}ggen},
  {Ogrean}, {Riseley}, {Cuciti}, {de Gasperin}, {Golovich}, {Kale}, {Venturi},
  {van Weeren}, {Wik}, \& {Wittman}}]{Bonafede17}
{Bonafede}, A., {Cassano}, R., {Br{\"u}ggen}, M., {et~al.} 2017, \mnras, 470,
  3465

\bibitem[{{Bonjean} {et~al.}(2018){Bonjean}, {Aghanim}, {Salom{\'e}},
  {Douspis}, \& {Beelen}}]{Bonjean2018}
{Bonjean}, V., {Aghanim}, N., {Salom{\'e}}, P., {Douspis}, M., \& {Beelen}, A.
  2018, \aap, 609, A49

\bibitem[{Botteon {et~al.}(2020)Botteon, van Weeren, Brunetti, de~Gasperin,
  Intema, Osinga, Di~Gennaro, Shimwell, Bonafede, Brüggen, Cassano, Cuciti,
  Dallacasa, Gastaldello, Mandal, Rossetti, \& Röttgering}]{Botteon2020}
Botteon, A., van Weeren, R.~J., Brunetti, G., {et~al.} 2020, Monthly Notices of
  the Royal Astronomical Society: Letters, 499, L11

\bibitem[{{Botteon} {et~al.}(2022){Botteon}, {van Weeren}, {Brunetti}, {Vazza},
  {Shimwell}, {Br{\"u}ggen}, {R{\"o}ttgering}, {de Gasperin}, {Akamatsu},
  {Bonafede}, {Cassano}, {Cuciti}, {Dallacasa}, {Di Gennaro}, \&
  {Gastaldello}}]{Botteon2022}
{Botteon}, A., {van Weeren}, R.~J., {Brunetti}, G., {et~al.} 2022, Science
  Advances, 8, eabq7623

\bibitem[{{Brentjens} \& {de Bruyn}(2005)}]{B&B05}
{Brentjens}, M.~A. \& {de Bruyn}, A.~G. 2005, \aap, 441, 1217

\bibitem[{{Brown} {et~al.}(2017){Brown}, {Vernstrom}, {Carretti}, {Dolag},
  {Gaensler}, {Staveley-Smith}, {Bernardi}, {Haverkorn}, {Kesteven}, \&
  {Poppi}}]{Brown17}
{Brown}, S., {Vernstrom}, T., {Carretti}, E., {et~al.} 2017, \mnras, 468, 4246

\bibitem[{{Brunetti} \& {Jones}(2014)}]{Brunetti2014}
{Brunetti}, G. \& {Jones}, T.~W. 2014, International Journal of Modern Physics
  D, 23, 1430007

\bibitem[{Brunetti \& Vazza(2020)}]{BrunettiVazza20}
Brunetti, G. \& Vazza, F. 2020, Phys. Rev. Lett., 124, 051101

\bibitem[{{Burn}(1966)}]{Burn1966}
{Burn}, B.~J. 1966, \mnras, 133, 67

\bibitem[{{Carretti} {et~al.}(2023){Carretti}, {O'Sullivan}, {Vacca}, {Vazza},
  {Gheller}, {Vernstrom}, \& {Bonafede}}]{Carretti2022b}
{Carretti}, E., {O'Sullivan}, S.~P., {Vacca}, V., {et~al.} 2023, \mnras, 518,
  2273

\bibitem[{{Carretti} {et~al.}(2022){Carretti}, {Vacca}, {O'Sullivan}, {Heald},
  {Horellou}, {R{\"o}ttgering}, {Scaife}, {Shimwell}, {Shulevski}, {Stuardi},
  \& {Vernstrom}}]{Carretti2022a}
{Carretti}, E., {Vacca}, V., {O'Sullivan}, S.~P., {et~al.} 2022, \mnras, 512,
  945

\bibitem[{{Cuciti} {et~al.}(2022){Cuciti}, {de Gasperin}, {Br{\"u}ggen},
  {Vazza}, {Brunetti}, {Shimwell}, {Edler}, {van Weeren}, {Botteon}, {Cassano},
  {Di Gennaro}, {Gastaldello}, {Drabent}, {R{\"o}ttgering}, \&
  {Tasse}}]{Cuciti2022}
{Cuciti}, V., {de Gasperin}, F., {Br{\"u}ggen}, M., {et~al.} 2022, \nat, 609,
  911

\bibitem[{{de Gasperin} {et~al.}(2019){de Gasperin}, {Dijkema}, {Drabent},
  {Mevius}, {Rafferty}, {van Weeren}, {Br{\"u}ggen}, {Callingham}, {Emig},
  {Heald}, {Intema}, {Morabito}, {Offringa}, {Oonk}, {Orr{\`u}},
  {R{\"o}ttgering}, {Sabater}, {Shimwell}, {Shulevski}, \&
  {Williams}}]{PREFACTOR2019}
{de Gasperin}, F., {Dijkema}, T.~J., {Drabent}, A., {et~al.} 2019, \aap, 622,
  A5

\bibitem[{{de Jong} {et~al.}(2022){de Jong}, {van Weeren}, {Botteon}, {Oonk},
  {Brunetti}, {Shimwell}, {Cassano}, {R{\"o}ttgering}, \& {Tasse}}]{deJong2022}
{de Jong}, J.~M.~G.~H.~J., {van Weeren}, R.~J., {Botteon}, A., {et~al.} 2022,
  \aap, 668, A107

\bibitem[{{Dom{\'\i}nguez-Fern{\'a}ndez}
  {et~al.}(2019){Dom{\'\i}nguez-Fern{\'a}ndez}, {Vazza}, {Br{\"u}ggen}, \&
  {Brunetti}}]{Dominguez-Fernandez2019}
{Dom{\'\i}nguez-Fern{\'a}ndez}, P., {Vazza}, F., {Br{\"u}ggen}, M., \&
  {Brunetti}, G. 2019, \mnras, 486, 623

\bibitem[{Donnert {et~al.}(2018)Donnert, Vazza, Br{\"u}ggen, \&
  ZuHone}]{Donnert2018}
Donnert, J., Vazza, F., Br{\"u}ggen, M., \& ZuHone, J. 2018, Space Science
  Reviews, 214, 122

\bibitem[{{Erceg} {et~al.}(2022){Erceg}, {Jeli{\'c}}, {Haverkorn}, {Bracco},
  {Shimwell}, {Tasse}, {Dickey}, {Ceraj}, {Drabent}, {Hardcastle}, \&
  {Turi{\'c}}}]{Erceg22}
{Erceg}, A., {Jeli{\'c}}, V., {Haverkorn}, M., {et~al.} 2022, \aap, 663, A7

\bibitem[{{Fujita} {et~al.}(1996){Fujita}, {Koyama}, {Tsuru}, \&
  {Matsumoto}}]{Fujita96}
{Fujita}, Y., {Koyama}, K., {Tsuru}, T., \& {Matsumoto}, H. 1996, \pasj, 48,
  191

\bibitem[{{Fujita} {et~al.}(2008){Fujita}, {Tawa}, {Hayashida}, {Takizawa},
  {Matsumoto}, {Okabe}, \& {Reiprich}}]{Fujita2008}
{Fujita}, Y., {Tawa}, N., {Hayashida}, K., {et~al.} 2008, \pasj, 60, S343

\bibitem[{{Govoni} \& {Feretti}(2004)}]{Govoni04}
{Govoni}, F. \& {Feretti}, L. 2004, International Journal of Modern Physics D,
  13, 1549

\bibitem[{Govoni {et~al.}(2019)Govoni, Orr{\`u}, Bonafede, Iacobelli, Paladino,
  Vazza, Murgia, Vacca, Giovannini, Feretti, Loi, Bernardi, Ferrari, Pizzo,
  Gheller, Manti, Br{\"u}ggen, Brunetti, Cassano, de~Gasperin, En{\ss}lin,
  Hoeft, Horellou, Junklewitz, R{\"o}ttgering, Scaife, Shimwell, van Weeren, \&
  Wise}]{Govoni19}
Govoni, F., Orr{\`u}, E., Bonafede, A., {et~al.} 2019, Science, 364, 981

\bibitem[{{Hincks} {et~al.}(2022){Hincks}, {Radiconi}, {Romero},
  {Madhavacheril}, {Mroczkowski}, {Austermann}, {Barbavara}, {Battaglia},
  {Battistelli}, {Bond}, {Calabrese}, {de Bernardis}, {Devlin}, {Dicker},
  {Duff}, {Duivenvoorden}, {Dunkley}, {D{\"u}nner}, {Gallardo}, {Govoni},
  {Hill}, {Hilton}, {Hubmayr}, {Hughes}, {Lamagna}, {Lokken}, {Masi}, {Mason},
  {McMahon}, {Moodley}, {Murgia}, {Naess}, {Page}, {Piacentini}, {Salatino},
  {Sarazin}, {Schillaci}, {Sievers}, {Sif{\'o}n}, {Staggs}, {Ullom}, {Vacca},
  {Van Engelen}, {Vissers}, {Wollack}, \& {Xu}}]{Hincks2021-bis}
{Hincks}, A.~D., {Radiconi}, F., {Romero}, C., {et~al.} 2022, \mnras, 510, 3335

\bibitem[{{Hoang} {et~al.}(2023){Hoang}, {Br{\"u}ggen}, {Zhang}, {Bonafede},
  {Liu}, {Liu}, {Shimwell}, {Botteon}, {Brunetti}, {Bulbul}, {Gennaro},
  {O'Sullivan}, {Pasini}, {R{\"o}ttgering}, {Vernstrom}, \& {van
  Weeren}}]{Hoang2023}
{Hoang}, D.~N., {Br{\"u}ggen}, M., {Zhang}, X., {et~al.} 2023, \mnras, 523,
  6320

\bibitem[{{Hodgson} {et~al.}(2022){Hodgson}, {Vazza}, {Johnston-Hollitt}, \&
  {McKinley}}]{Hodgson2021}
{Hodgson}, T., {Vazza}, F., {Johnston-Hollitt}, M., \& {McKinley}, B. 2022,
  \pasa, 39, e033

\bibitem[{{Locatelli} {et~al.}(2021){Locatelli}, {Vazza}, {Bonafede}, {Banfi},
  {Bernardi}, {Gheller}, {Botteon}, \& {Shimwell}}]{Locatelli2021}
{Locatelli}, N., {Vazza}, F., {Bonafede}, A., {et~al.} 2021, \aap, 652, A80

\bibitem[{{Mohan} \& {Rafferty}(2015)}]{pybdsf}
{Mohan}, N. \& {Rafferty}, D. 2015, {PyBDSF: Python Blob Detection and Source
  Finder}, Astrophysics Source Code Library, record ascl:1502.007

\bibitem[{Murgia {et~al.}(2010)Murgia, Govoni, Feretti, \&
  Giovannini}]{Murgia2010}
Murgia, M., Govoni, F., Feretti, L., \& Giovannini, G. 2010, Astronomy and
  Astrophysics, 509, 1

\bibitem[{{Natwariya}(2021)}]{Natwariya21}
{Natwariya}, P.~K. 2021, European Physical Journal C, 81, 394

\bibitem[{{Neronov} \& {Vovk}(2010)}]{Neronov2010}
{Neronov}, A. \& {Vovk}, I. 2010, Science, 328, 73

\bibitem[{{Nunhokee} {et~al.}(2023){Nunhokee}, {Bernardi}, {Manti}, {Govoni},
  {Bonafede}, {Venturi}, {Dallacasa}, {Murgia}, {Pizzo}, {Smirnov}, \&
  {Vacca}}]{Nunhokee2023}
{Nunhokee}, C.~D., {Bernardi}, G., {Manti}, S., {et~al.} 2023, \mnras, 522,
  4421

\bibitem[{{Offringa} {et~al.}(2014){Offringa}, {McKinley}, {Hurley-Walker},
  {Briggs}, {Wayth}, {Kaplan}, {Bell}, {Feng}, {Neben}, {Hughes}, {Rhee},
  {Murphy}, {Bhat}, {Bernardi}, {Bowman}, {Cappallo}, {Corey}, {Deshpande},
  {Emrich}, {Ewall-Wice}, {Gaensler}, {Goeke}, {Greenhill}, {Hazelton},
  {Hindson}, {Johnston-Hollitt}, {Jacobs}, {Kasper}, {Kratzenberg}, {Lenc},
  {Lonsdale}, {Lynch}, {McWhirter}, {Mitchell}, {Morales}, {Morgan},
  {Kudryavtseva}, {Oberoi}, {Ord}, {Pindor}, {Procopio}, {Prabu}, {Riding},
  {Roshi}, {Shankar}, {Srivani}, {Subrahmanyan}, {Tingay}, {Waterson},
  {Webster}, {Whitney}, {Williams}, \& {Williams}}]{Wsclean14}
{Offringa}, A.~R., {McKinley}, B., {Hurley-Walker}, N., {et~al.} 2014, \mnras,
  444, 606

\bibitem[{{Offringa} \& {Smirnov}(2017)}]{Offringa2017}
{Offringa}, A.~R. \& {Smirnov}, O. 2017, \mnras, 471, 301

\bibitem[{{O'Sullivan} {et~al.}(2012){O'Sullivan}, {Brown}, {Robishaw},
  {Schnitzeler}, {McClure-Griffiths}, {Feain}, {Taylor}, {Gaensler},
  {Landecker}, {Harvey-Smith}, \& {Carretti}}]{OSull2012}
{O'Sullivan}, S.~P., {Brown}, S., {Robishaw}, T., {et~al.} 2012, \mnras, 421,
  3300

\bibitem[{{O'Sullivan} {et~al.}(2020){O'Sullivan}, {Br{\"u}ggen}, {Vazza},
  {Carretti}, {Locatelli}, {Stuardi}, {Vacca}, {Vernstrom}, {Heald},
  {Horellou}, {Shimwell}, {Hardcastle}, {Tasse}, \&
  {R{\"o}ttgering}}]{OSullivan2020}
{O'Sullivan}, S.~P., {Br{\"u}ggen}, M., {Vazza}, F., {et~al.} 2020, \mnras,
  495, 2607

\bibitem[{{O'Sullivan} {et~al.}(2019){O'Sullivan}, {Machalski}, {Van Eck},
  {Heald}, {Br{\"u}ggen}, {Fynbo}, {Heintz}, {Lara-Lopez}, {Vacca},
  {Hardcastle}, {Shimwell}, {Tasse}, {Vazza}, {Andernach}, {Birkinshaw},
  {Haverkorn}, {Horellou}, {Williams}, {Harwood}, {Brunetti}, {Anderson},
  {Mao}, {Nikiel-Wroczy{\'n}ski}, {Takahashi}, {Carretti}, {Vernstrom}, {van
  Weeren}, {Orr{\'u}}, {Morabito}, \& {Callingham}}]{OSullivan2019}
{O'Sullivan}, S.~P., {Machalski}, J., {Van Eck}, C.~L., {et~al.} 2019, \aap,
  622, A16

\bibitem[{{O'Sullivan} {et~al.}(2023){O'Sullivan}, {Shimwell}, {Hardcastle},
  {Tasse}, {Heald}, {Carretti}, {Br{\"u}ggen}, {Vacca}, {Sobey}, {Van Eck},
  {Horellou}, {Beck}, {Bilicki}, {Bourke}, {Botteon}, {Croston}, {Drabent},
  {Duncan}, {Heesen}, {Ideguchi}, {Kirwan}, {Lawlor}, {Mingo},
  {Nikiel-Wroczy{\'n}ski}, {Piotrowska}, {Scaife}, \& {van
  Weeren}}]{OSullivan2023}
{O'Sullivan}, S.~P., {Shimwell}, T.~W., {Hardcastle}, M.~J., {et~al.} 2023,
  \mnras, 519, 5723

\bibitem[{{Paoletti} \& {Finelli}(2019)}]{Paoletti19}
{Paoletti}, D. \& {Finelli}, F. 2019, \jcap, 2019, 028

\bibitem[{{Planck Collaboration} {et~al.}(2013){Planck Collaboration}, {Ade},
  {Aghanim}, {Arnaud}, {Ashdown}, {Atrio-Barandela}, {Aumont}, {Baccigalupi},
  {Balbi}, {Banday}, {Barreiro}, {Battaner}, {Benabed}, {Beno{\^\i}t},
  {Bernard}, {Bersanelli}, {Bhatia}, {Bikmaev}, {B{\"o}hringer}, {Bonaldi},
  {Bond}, {Borrill}, {Bouchet}, {Bourdin}, {Burenin}, {Burigana}, {Cabella},
  {Cardoso}, {Castex}, {Catalano}, {Cay{\'o}n}, {Chamballu}, {Chary}, {Chiang},
  {Chon}, {Christensen}, {Clements}, {Colafrancesco}, {Colombo}, {Comis},
  {Coulais}, {Crill}, {Cuttaia}, {Danese}, {Davis}, {de Bernardis}, {de
  Gasperis}, {de Zotti}, {Delabrouille}, {D{\'e}mocl{\`e}s}, {D{\'e}sert},
  {Diego}, {Dolag}, {Dole}, {Donzelli}, {Dor{\'e}}, {D{\"o}rl}, {Douspis},
  {Dupac}, {Efstathiou}, {En{\ss}lin}, {Eriksen}, {Finelli}, {Flores-Cacho},
  {Forni}, {Frailis}, {Franceschi}, {Frommert}, {Ganga}, {G{\'e}nova-Santos},
  {Giard}, {Gilfanov}, {Giraud-H{\'e}raud}, {Gonz{\'a}lez-Nuevo}, {G{\'o}rski},
  {Gregorio}, {Gruppuso}, {Hansen}, {Harrison}, {Hempel},
  {Henrot-Versill{\'e}}, {Hern{\'a}ndez-Monteagudo}, {Herranz}, {Hildebrandt},
  {Hivon}, {Hobson}, {Holmes}, {Hovest}, {Hurier}, {Jaffe}, {Jaffe},
  {Jagemann}, {Jones}, {Juvela}, {Khamitov}, {Kisner}, {Kneissl}, {Knoche},
  {Knox}, {Kunz}, {Kurki-Suonio}, {Lagache}, {Lamarre}, {Lasenby}, {Lawrence},
  {Le Jeune}, {Leonardi}, {Lilje}, {Linden-V{\o}rnle}, {L{\'o}pez-Caniego},
  {Lubin}, {Luzzi}, {Mac{\'\i}as-P{\'e}rez}, {Maffei}, {Maino}, {Mandolesi},
  {Maris}, {Marleau}, {Marshall}, {Mart{\'\i}nez-Gonz{\'a}lez}, {Masi},
  {Massardi}, {Matarrese}, {Matthai}, {Mazzotta}, {Mei}, {Melchiorri}, {Melin},
  {Mendes}, {Mennella}, {Mitra}, {Miville-Desch{\`e}nes}, {Moneti}, {Montier},
  {Morgante}, {Munshi}, {Murphy}, {Naselsky}, {Nati}, {Natoli},
  {N{\o}rgaard-Nielsen}, {Noviello}, {Novikov}, {Novikov}, {Osborne}, {Pajot},
  {Paoletti}, {Pasian}, {Patanchon}, {Perdereau}, {Perotto}, {Perrotta},
  {Piacentini}, {Piat}, {Pierpaoli}, {Piffaretti}, {Plaszczynski},
  {Pointecouteau}, {Polenta}, {Ponthieu}, {Popa}, {Poutanen}, {Pratt},
  {Prunet}, {Puget}, {Rachen}, {Rebolo}, {Reinecke}, {Remazeilles}, {Renault},
  {Ricciardi}, {Riller}, {Ristorcelli}, {Rocha}, {Roman}, {Rosset}, {Rossetti},
  {Rubi{\~n}o-Mart{\'\i}n}, {Rusholme}, {Sandri}, {Savini}, {Schaefer},
  {Scott}, {Smoot}, {Starck}, {Sudiwala}, {Sunyaev}, {Sutton}, {Suur-Uski},
  {Sygnet}, {Tauber}, {Terenzi}, {Toffolatti}, {Tomasi}, {Tristram}, {Tucci},
  {Valenziano}, {Van Tent}, {Vielva}, {Villa}, {Vittorio}, {Wade}, {Wandelt},
  {Welikala}, {White}, {Yvon}, {Zacchei}, \& {Zonca}}]{Planck2013}
{Planck Collaboration}, {Ade}, P.~A.~R., {Aghanim}, N., {et~al.} 2013, \aap,
  550, A134

\bibitem[{{Planck Collaboration} {et~al.}(2020){Planck Collaboration},
  {Aghanim}, {Akrami}, {Ashdown}, {Aumont}, {Baccigalupi}, {Ballardini},
  {Banday}, {Barreiro}, {Bartolo}, {Basak}, {Battye}, {Benabed}, {Bernard},
  {Bersanelli}, {Bielewicz}, {Bock}, {Bond}, {Borrill}, {Bouchet}, {Boulanger},
  {Bucher}, {Burigana}, {Butler}, {Calabrese}, {Cardoso}, {Carron},
  {Challinor}, {Chiang}, {Chluba}, {Colombo}, {Combet}, {Contreras}, {Crill},
  {Cuttaia}, {de Bernardis}, {de Zotti}, {Delabrouille}, {Delouis}, {Di
  Valentino}, {Diego}, {Dor{\'e}}, {Douspis}, {Ducout}, {Dupac}, {Dusini},
  {Efstathiou}, {Elsner}, {En{\ss}lin}, {Eriksen}, {Fantaye}, {Farhang},
  {Fergusson}, {Fernandez-Cobos}, {Finelli}, {Forastieri}, {Frailis},
  {Fraisse}, {Franceschi}, {Frolov}, {Galeotta}, {Galli}, {Ganga},
  {G{\'e}nova-Santos}, {Gerbino}, {Ghosh}, {Gonz{\'a}lez-Nuevo}, {G{\'o}rski},
  {Gratton}, {Gruppuso}, {Gudmundsson}, {Hamann}, {Handley}, {Hansen},
  {Herranz}, {Hildebrandt}, {Hivon}, {Huang}, {Jaffe}, {Jones}, {Karakci},
  {Keih{\"a}nen}, {Keskitalo}, {Kiiveri}, {Kim}, {Kisner}, {Knox},
  {Krachmalnicoff}, {Kunz}, {Kurki-Suonio}, {Lagache}, {Lamarre}, {Lasenby},
  {Lattanzi}, {Lawrence}, {Le Jeune}, {Lemos}, {Lesgourgues}, {Levrier},
  {Lewis}, {Liguori}, {Lilje}, {Lilley}, {Lindholm}, {L{\'o}pez-Caniego},
  {Lubin}, {Ma}, {Mac{\'\i}as-P{\'e}rez}, {Maggio}, {Maino}, {Mandolesi},
  {Mangilli}, {Marcos-Caballero}, {Maris}, {Martin}, {Martinelli},
  {Mart{\'\i}nez-Gonz{\'a}lez}, {Matarrese}, {Mauri}, {McEwen}, {Meinhold},
  {Melchiorri}, {Mennella}, {Migliaccio}, {Millea}, {Mitra},
  {Miville-Desch{\^e}nes}, {Molinari}, {Montier}, {Morgante}, {Moss}, {Natoli},
  {N{\o}rgaard-Nielsen}, {Pagano}, {Paoletti}, {Partridge}, {Patanchon},
  {Peiris}, {Perrotta}, {Pettorino}, {Piacentini}, {Polastri}, {Polenta},
  {Puget}, {Rachen}, {Reinecke}, {Remazeilles}, {Renzi}, {Rocha}, {Rosset},
  {Roudier}, {Rubi{\~n}o-Mart{\'\i}n}, {Ruiz-Granados}, {Salvati}, {Sandri},
  {Savelainen}, {Scott}, {Shellard}, {Sirignano}, {Sirri}, {Spencer},
  {Sunyaev}, {Suur-Uski}, {Tauber}, {Tavagnacco}, {Tenti}, {Toffolatti},
  {Tomasi}, {Trombetti}, {Valenziano}, {Valiviita}, {Van Tent}, {Vibert},
  {Vielva}, {Villa}, {Vittorio}, {Wandelt}, {Wehus}, {White}, {White},
  {Zacchei}, \& {Zonca}}]{Planck20}
{Planck Collaboration}, {Aghanim}, N., {Akrami}, Y., {et~al.} 2020, \aap, 641,
  A6

\bibitem[{{Pomakov} {et~al.}(2022){Pomakov}, {O'Sullivan}, {Br{\"u}ggen},
  {Vazza}, {Carretti}, {Heald}, {Horellou}, {Shimwell}, {Shulevski}, \&
  {Vernstrom}}]{Pomakov2022}
{Pomakov}, V.~P., {O'Sullivan}, S.~P., {Br{\"u}ggen}, M., {et~al.} 2022,
  \mnras, 515, 256

\bibitem[{{Pshirkov} {et~al.}(2016){Pshirkov}, {Tinyakov}, \&
  {Urban}}]{Pshirkov16}
{Pshirkov}, M.~S., {Tinyakov}, P.~G., \& {Urban}, F.~R. 2016, \prl, 116, 191302

\bibitem[{{Purcell} {et~al.}(2020){Purcell}, {Van Eck}, {West}, {Sun}, \&
  {Gaensler}}]{RMsynth}
{Purcell}, C.~R., {Van Eck}, C.~L., {West}, J., {Sun}, X.~H., \& {Gaensler},
  B.~M. 2020, {RM-Tools: Rotation measure (RM) synthesis and Stokes QU-fitting}

\bibitem[{{Radiconi} {et~al.}(2022){Radiconi}, {Vacca}, {Battistelli},
  {Bonafede}, {Capalbo}, {Devlin}, {Di Mascolo}, {Feretti}, {Gallardo}, {Gill},
  {Giovannini}, {Govoni}, {Guan}, {Hilton}, {Hincks}, {Hughes}, {Iacobelli},
  {Isopi}, {Loi}, {Moodley}, {Mroczkowski}, {Murgia}, {Orr{\'u}}, {Paladino},
  {Partridge}, {Sarazin}, {Orlowski Scherer}, {Sif{\'o}n}, {Vargas}, {Vazza},
  \& {Wollack}}]{Radiconi2022}
{Radiconi}, F., {Vacca}, V., {Battistelli}, E., {et~al.} 2022, \mnras, 517,
  5232

\bibitem[{{Rajpurohit} {et~al.}(2021{\natexlab{a}}){Rajpurohit}, {Brunetti},
  {Bonafede}, {van Weeren}, {Botteon}, {Vazza}, {Hoeft}, {Riseley},
  {Bonnassieux}, {Brienza}, {Forman}, {R{\"o}ttgering}, {Rajpurohit},
  {Locatelli}, {Shimwell}, {Cassano}, {Di Gennaro}, {Br{\"u}ggen}, {Wittor},
  {Drabent}, \& {Ignesti}}]{Kamlesh2021-j0717}
{Rajpurohit}, K., {Brunetti}, G., {Bonafede}, A., {et~al.} 2021{\natexlab{a}},
  \aap, 646, A135

\bibitem[{{Rajpurohit} {et~al.}(2021{\natexlab{b}}){Rajpurohit}, {Vazza}, {van
  Weeren}, {Hoeft}, {Brienza}, {Bonnassieux}, {Riseley}, {Brunetti},
  {Bonafede}, {Br{\"u}ggen}, {Formann}, {Rajpurohit}, {R{\"o}ttgering},
  {Drabent}, {Dom{\'\i}nguez-Fern{\'a}ndez}, {Wittor}, \&
  {Andrade-Santos}}]{Kamlesh2021-A2744}
{Rajpurohit}, K., {Vazza}, F., {van Weeren}, R.~J., {et~al.}
  2021{\natexlab{b}}, \aap, 654, A41

\bibitem[{{Sakelliou} \& {Ponman}(2004)}]{Sakelliou04}
{Sakelliou}, I. \& {Ponman}, T.~J. 2004, \mnras, 351, 1439

\bibitem[{Shimwell {et~al.}(2022)Shimwell, Hardcastle, Tasse, Best,
  R{\"{o}}ttgering, Williams, Botteon, Drabent, Mechev, Shulevski, van Weeren,
  \& Al.}]{Shimwell2022}
Shimwell, T.~W., Hardcastle, M.~J., Tasse, C., {et~al.} 2022, Astronomy \&
  Astrophysics, 1

\bibitem[{{Shimwell} {et~al.}(2017){Shimwell}, {R{\"o}ttgering}, {Best},
  {Williams}, {Dijkema}, {de Gasperin}, {Hardcastle}, {Heald}, {Hoang},
  {Horneffer}, {Intema}, {Mahony}, {Mandal}, {Mechev}, {Morabito}, {Oonk},
  {Rafferty}, {Retana-Montenegro}, {Sabater}, {Tasse}, {van Weeren},
  {Br{\"u}ggen}, {Brunetti}, {Chy{\.z}y}, {Conway}, {Haverkorn}, {Jackson},
  {Jarvis}, {McKean}, {Miley}, {Morganti}, {White}, {Wise}, {van Bemmel},
  {Beck}, {Brienza}, {Bonafede}, {Calistro Rivera}, {Cassano}, {Clarke},
  {Cseh}, {Deller}, {Drabent}, {van Driel}, {Engels}, {Falcke}, {Ferrari},
  {Fr{\"o}hlich}, {Garrett}, {Harwood}, {Heesen}, {Hoeft}, {Horellou},
  {Israel}, {Kapi{\'n}ska}, {Kunert-Bajraszewska}, {McKay}, {Mohan},
  {Orr{\'u}}, {Pizzo}, {Prandoni}, {Schwarz}, {Shulevski}, {Sipior}, {Smith},
  {Sridhar}, {Steinmetz}, {Stroe}, {Varenius}, {van der Werf}, {Zensus}, \&
  {Zwart}}]{LoTSS2017}
{Shimwell}, T.~W., {R{\"o}ttgering}, H.~J.~A., {Best}, P.~N., {et~al.} 2017,
  \aap, 598, A104

\bibitem[{{Shimwell} {et~al.}(2019){Shimwell}, {Tasse}, {Hardcastle}, {Mechev},
  {Williams}, {Best}, {R{\"o}ttgering}, {Callingham}, {Dijkema}, {de Gasperin},
  {Hoang}, {Hugo}, {Mirmont}, {Oonk}, {Prandoni}, {Rafferty}, {Sabater},
  {Smirnov}, {van Weeren}, {White}, {Atemkeng}, {Bester}, {Bonnassieux},
  {Br{\"u}ggen}, {Brunetti}, {Chy{\.z}y}, {Cochrane}, {Conway}, {Croston},
  {Danezi}, {Duncan}, {Haverkorn}, {Heald}, {Iacobelli}, {Intema}, {Jackson},
  {Jamrozy}, {Jarvis}, {Lakhoo}, {Mevius}, {Miley}, {Morabito}, {Morganti},
  {Nisbet}, {Orr{\'u}}, {Perkins}, {Pizzo}, {Schrijvers}, {Smith}, {Vermeulen},
  {Wise}, {Alegre}, {Bacon}, {van Bemmel}, {Beswick}, {Bonafede}, {Botteon},
  {Bourke}, {Brienza}, {Calistro Rivera}, {Cassano}, {Clarke}, {Conselice},
  {Dettmar}, {Drabent}, {Dumba}, {Emig}, {En{\ss}lin}, {Ferrari}, {Garrett},
  {G{\'e}nova-Santos}, {Goyal}, {G{\"u}rkan}, {Hale}, {Harwood}, {Heesen},
  {Hoeft}, {Horellou}, {Jackson}, {Kokotanekov}, {Kondapally},
  {Kunert-Bajraszewska}, {Mahatma}, {Mahony}, {Mandal}, {McKean}, {Merloni},
  {Mingo}, {Miskolczi}, {Mooney}, {Nikiel-Wroczy{\'n}ski}, {O'Sullivan},
  {Quinn}, {Reich}, {Roskowi{\'n}ski}, {Rowlinson}, {Savini}, {Saxena},
  {Schwarz}, {Shulevski}, {Sridhar}, {Stacey}, {Urquhart}, {van der Wiel},
  {Varenius}, {Webster}, \& {Wilber}}]{LoTSS-DR1}
{Shimwell}, T.~W., {Tasse}, C., {Hardcastle}, M.~J., {et~al.} 2019, \aap, 622,
  A1

\bibitem[{{Sokoloff} {et~al.}(1998){Sokoloff}, {Bykov}, {Shukurov},
  {Berkhuijsen}, {Beck}, \& {Poezd}}]{Sokoloff1998}
{Sokoloff}, D.~D., {Bykov}, A.~A., {Shukurov}, A., {et~al.} 1998, \mnras, 299,
  189

\bibitem[{{Subramanian}(2016)}]{Subramanian2016}
{Subramanian}, K. 2016, Reports on Progress in Physics, 79, 076901

\bibitem[{{Sunyaev} \& {Zeldovich}(1969)}]{SZ1969}
{Sunyaev}, R.~A. \& {Zeldovich}, Y.~B. 1969, \nat, 223, 721

\bibitem[{{Tribble}(1991)}]{Tribble1991}
{Tribble}, P.~C. 1991, \mnras, 250, 726

\bibitem[{{Van Eck} {et~al.}(2019){Van Eck}, {Haverkorn}, {Alves}, {Beck},
  {Best}, {Carretti}, {Chy{\.z}y}, {En{\ss}lin}, {Farnes}, {Ferri{\`e}re},
  {Heald}, {Iacobelli}, {Jeli{\'c}}, {Reich}, {R{\"o}ttgering}, \&
  {Schnitzeler}}]{Van_Eck19}
{Van Eck}, C.~L., {Haverkorn}, M., {Alves}, M.~I.~R., {et~al.} 2019, \aap, 623,
  A71

\bibitem[{{van Haarlem} {et~al.}(2013){van Haarlem}, {Wise}, {Gunst}, {Heald},
  {McKean}, {Hessels}, {de Bruyn}, {Nijboer}, {Swinbank}, {Fallows},
  {Brentjens}, {Nelles}, {Beck}, {Falcke}, {Fender}, {H{\"o}randel},
  {Koopmans}, {Mann}, {Miley}, {R{\"o}ttgering}, {Stappers}, {Wijers},
  {Zaroubi}, {van den Akker}, {Alexov}, {Anderson}, {Anderson}, {van Ardenne},
  {Arts}, {Asgekar}, {Avruch}, {Batejat}, {B{\"a}hren}, {Bell}, {Bell}, {van
  Bemmel}, {Bennema}, {Bentum}, {Bernardi}, {Best}, {B{\^\i}rzan}, {Bonafede},
  {Boonstra}, {Braun}, {Bregman}, {Breitling}, {van de Brink}, {Broderick},
  {Broekema}, {Brouw}, {Br{\"u}ggen}, {Butcher}, {van Cappellen}, {Ciardi},
  {Coenen}, {Conway}, {Coolen}, {Corstanje}, {Damstra}, {Davies}, {Deller},
  {Dettmar}, {van Diepen}, {Dijkstra}, {Donker}, {Doorduin}, {Dromer}, {Drost},
  {van Duin}, {Eisl{\"o}ffel}, {van Enst}, {Ferrari}, {Frieswijk}, {Gankema},
  {Garrett}, {de Gasperin}, {Gerbers}, {de Geus}, {Grie{\ss}meier}, {Grit},
  {Gruppen}, {Hamaker}, {Hassall}, {Hoeft}, {Holties}, {Horneffer}, {van der
  Horst}, {van Houwelingen}, {Huijgen}, {Iacobelli}, {Intema}, {Jackson},
  {Jelic}, {de Jong}, {Juette}, {Kant}, {Karastergiou}, {Koers}, {Kollen},
  {Kondratiev}, {Kooistra}, {Koopman}, {Koster}, {Kuniyoshi}, {Kramer},
  {Kuper}, {Lambropoulos}, {Law}, {van Leeuwen}, {Lemaitre}, {Loose}, {Maat},
  {Macario}, {Markoff}, {Masters}, {McFadden}, {McKay-Bukowski}, {Meijering},
  {Meulman}, {Mevius}, {Middelberg}, {Millenaar}, {Miller-Jones}, {Mohan},
  {Mol}, {Morawietz}, {Morganti}, {Mulcahy}, {Mulder}, {Munk}, {Nieuwenhuis},
  {van Nieuwpoort}, {Noordam}, {Norden}, {Noutsos}, {Offringa}, {Olofsson},
  {Omar}, {Orr{\'u}}, {Overeem}, {Paas}, {Pandey-Pommier}, {Pandey}, {Pizzo},
  {Polatidis}, {Rafferty}, {Rawlings}, {Reich}, {de Reijer}, {Reitsma},
  {Renting}, {Riemers}, {Rol}, {Romein}, {Roosjen}, {Ruiter}, {Scaife}, {van
  der Schaaf}, {Scheers}, {Schellart}, {Schoenmakers}, {Schoonderbeek},
  {Serylak}, {Shulevski}, {Sluman}, {Smirnov}, {Sobey}, {Spreeuw}, {Steinmetz},
  {Sterks}, {Stiepel}, {Stuurwold}, {Tagger}, {Tang}, {Tasse}, {Thomas},
  {Thoudam}, {Toribio}, {van der Tol}, {Usov}, {van Veelen}, {van der Veen},
  {ter Veen}, {Verbiest}, {Vermeulen}, {Vermaas}, {Vocks}, {Vogt}, {de Vos},
  {van der Wal}, {van Weeren}, {Weggemans}, {Weltevrede}, {White}, {Wijnholds},
  {Wilhelmsson}, {Wucknitz}, {Yatawatta}, {Zarka}, {Zensus}, \& {van
  Zwieten}}]{LOFAR2013}
{van Haarlem}, M.~P., {Wise}, M.~W., {Gunst}, A.~W., {et~al.} 2013, \aap, 556,
  A2

\bibitem[{van Weeren {et~al.}(2019)van Weeren, de~Gasperin, Akamatsu,
  Br{\"u}ggen, Feretti, Kang, Stroe, \& Zandanel}]{vanWeeren19}
van Weeren, R.~J., de~Gasperin, F., Akamatsu, H., {et~al.} 2019, Space Science
  Reviews, 215, 16

\bibitem[{{van Weeren} {et~al.}(2021){van Weeren}, {Shimwell}, {Botteon},
  {Brunetti}, {Br{\"u}ggen}, {Boxelaar}, {Cassano}, {Di Gennaro},
  {Andrade-Santos}, {Bonnassieux}, {Bonafede}, {Cuciti}, {Dallacasa}, {de
  Gasperin}, {Gastaldello}, {Hardcastle}, {Hoeft}, {Kraft}, {Mandal},
  {Rossetti}, {R{\"o}ttgering}, {Tasse}, \& {Wilber}}]{vanWeeren2021}
{van Weeren}, R.~J., {Shimwell}, T.~W., {Botteon}, A., {et~al.} 2021, \aap,
  651, A115

\bibitem[{{Vazza} {et~al.}(2017){Vazza}, {Br{\"u}ggen}, {Gheller}, {Hackstein},
  {Wittor}, \& {Hinz}}]{Vazza2017}
{Vazza}, F., {Br{\"u}ggen}, M., {Gheller}, C., {et~al.} 2017, Classical and
  Quantum Gravity, 34, 234001

\bibitem[{{Vazza} {et~al.}(2018){Vazza}, {Brunetti}, {Br{\"u}ggen}, \&
  {Bonafede}}]{Vazza2018}
{Vazza}, F., {Brunetti}, G., {Br{\"u}ggen}, M., \& {Bonafede}, A. 2018, \mnras,
  474, 1672

\bibitem[{{Vazza} {et~al.}(2009){Vazza}, {Brunetti}, \& {Gheller}}]{Vazza09}
{Vazza}, F., {Brunetti}, G., \& {Gheller}, C. 2009, \mnras, 395, 1333

\bibitem[{{Vazza} {et~al.}(2021){Vazza}, Locatelli, Rajpurohit, Banfi,
  Dom{\'{i}}nguez-Fern{\'{a}}ndez, Wittor, Angelinelli, Inchingolo, Brienza,
  Hackstein, Dallacasa, Gheller, Br{\"{u}}ggen, Brunetti, Bonafede, Ettori,
  Stuardi, Paoletti, \& Finelli}]{Vazza2021}
{Vazza}, F., Locatelli, N., Rajpurohit, K., {et~al.} 2021, Galaxies, 9, 109

\bibitem[{{Venturi} {et~al.}(2008){Venturi}, {Giacintucci}, {Dallacasa},
  {Cassano}, {Brunetti}, {Bardelli}, \& {Setti}}]{Venturi08}
{Venturi}, T., {Giacintucci}, S., {Dallacasa}, D., {et~al.} 2008, \aap, 484,
  327

\bibitem[{{Venturi} {et~al.}(2022){Venturi}, {Giacintucci}, {Merluzzi},
  {Bardelli}, {Busarello}, {Dallacasa}, {Sikhosana}, {Marvil}, {Smirnov},
  {Bourdin}, {Mazzotta}, {Rossetti}, {Rudnick}, {Bernardi}, {Br{\"u}ggen},
  {Carretti}, {Cassano}, {Di Gennaro}, {Gastaldello}, {Kale}, {Knowles},
  {Koribalski}, {Heywood}, {Hopkins}, {Norris}, {Reiprich}, {Tasse},
  {Vernstrom}, {Zucca}, {Bester}, {Diego}, \& {Kanapathippillai}}]{Venturi2022}
{Venturi}, T., {Giacintucci}, S., {Merluzzi}, P., {et~al.} 2022, \aap, 660, A81

\bibitem[{{Vernstrom} {et~al.}(2017){Vernstrom}, {Gaensler}, {Brown}, {Lenc},
  \& {Norris}}]{Vernstrom17}
{Vernstrom}, T., {Gaensler}, B.~M., {Brown}, S., {Lenc}, E., \& {Norris}, R.~P.
  2017, \mnras, 467, 4914

\bibitem[{{Vernstrom} {et~al.}(2019){Vernstrom}, {Gaensler}, {Rudnick}, \&
  {Andernach}}]{Vernstrom19}
{Vernstrom}, T., {Gaensler}, B.~M., {Rudnick}, L., \& {Andernach}, H. 2019,
  \apj, 878, 92

\bibitem[{{Vernstrom} {et~al.}(2018){Vernstrom}, {Gaensler}, {Vacca}, {Farnes},
  {Haverkorn}, \& {O'Sullivan}}]{Vernstrom2018}
{Vernstrom}, T., {Gaensler}, B.~M., {Vacca}, V., {et~al.} 2018, \mnras, 475,
  1736

\bibitem[{{Vernstrom} {et~al.}(2021){Vernstrom}, {Heald}, {Vazza}, {Galvin},
  {West}, {Locatelli}, {Fornengo}, \& {Pinetti}}]{Vernstrom2021}
{Vernstrom}, T., {Heald}, G., {Vazza}, F., {et~al.} 2021, \mnras, 505, 4178

\bibitem[{{Vernstrom} {et~al.}(2023){Vernstrom}, {West}, {Vazza}, {Wittor},
  {Riseley}, \& {Heald}}]{Vernstrom2023}
{Vernstrom}, T., {West}, J., {Vazza}, F., {et~al.} 2023, Science Advances, 9,
  eade7233

\bibitem[{{Widrow} {et~al.}(2012){Widrow}, {Ryu}, {Schleicher}, {Subramanian},
  {Tsagas}, \& {Treumann}}]{Widrow2012}
{Widrow}, L.~M., {Ryu}, D., {Schleicher}, D. R.~G., {et~al.} 2012, \ssr, 166,
  37

\bibitem[{{Wittor} {et~al.}(2019{\natexlab{a}}){Wittor},
  {Dom{\'\i}nguez-Fern{\'a}ndez}, {Vazza}, \& {Br{\"u}ggen}}]{Wittor2019}
{Wittor}, D., {Dom{\'\i}nguez-Fern{\'a}ndez}, P., {Vazza}, F., \&
  {Br{\"u}ggen}, M. 2019{\natexlab{a}}, arXiv e-prints, arXiv:1909.10792

\bibitem[{{Wittor} {et~al.}(2019{\natexlab{b}}){Wittor}, {Hoeft}, {Vazza},
  {Br{\"u}ggen}, \& {Dom{\'\i}nguez-Fern{\'a}ndez}}]{WittorRR19}
{Wittor}, D., {Hoeft}, M., {Vazza}, F., {Br{\"u}ggen}, M., \&
  {Dom{\'\i}nguez-Fern{\'a}ndez}, P. 2019{\natexlab{b}}, \mnras, 490, 3987

\end{thebibliography}
%

\end{document}